\documentclass[apj]{emulateapj}
\usepackage{mathrsfs}

\newcommand{\kms}         {km~s$^{-1}$}

\newcommand{\ha}          {\mbox{H$\alpha$}}

\shorttitle{Complete Spectroscopic Survey of Segue 1}
\shortauthors{Simon et al.}

\begin{document}

\title{A Complete Spectroscopic Survey of the Milky Way Satellite
  Segue~1: The Darkest Galaxy\footnotemark[*]}

\author{Joshua D. Simon\altaffilmark{1}, Marla Geha\altaffilmark{2},
  Quinn E. Minor\altaffilmark{3}, Gregory D. Martinez\altaffilmark{3},
  Evan N. Kirby\altaffilmark{4,5}, James S. Bullock\altaffilmark{3},
  Manoj Kaplinghat\altaffilmark{3}, Louis
  E. Strigari\altaffilmark{6,5}, Beth Willman\altaffilmark{7}, Philip
  I. Choi\altaffilmark{8}, Erik J. Tollerud\altaffilmark{3}, and Joe
  Wolf\altaffilmark{3}}

\footnotetext[*]{The data presented herein were obtained at the
  W. M. Keck Observatory, which is operated as a scientific
  partnership among the California Institute of Technology, the
  University of California, and NASA.  The Observatory was made
  possible by the generous financial support of the W. M. Keck
  Foundation.}

\altaffiltext{1}{Observatories of the Carnegie Institution of
  Washington, 813 Santa Barbara St., Pasadena, CA 91101;
  jsimon@obs.carnegiescience.edu}

\altaffiltext{2}{Astronomy Department, Yale University, New Haven,
  CT~06520; marla.geha@yale.edu}

\altaffiltext{3}{Center for Cosmology, Department of Physics and
  Astronomy, University of California, Irvine, CA 92697;
  qminor@uci.edu, gmartine@uci.edu, bullock@uci.edu, mkapling@uci.edu,
  etolleru@uci.edu, wolfj@uci.edu}

\altaffiltext{4}{California Institute of Technology, Department of
  Astronomy, MS 249-17, Pasadena, CA  91106; enk@astro.caltech.edu}

\altaffiltext{5}{Hubble Fellow}

\altaffiltext{6}{Kavli Institute for Particle Astrophysics and
  Cosmology, Stanford University, Stanford, CA 94305;
  strigari@stanford.edu}

\altaffiltext{7}{Departments of Physics and Astronomy, Haverford
  College, Haverford, PA  19041; bwillman@haverford.edu}

\altaffiltext{8}{Department of Physics and Astronomy, Pomona
  College, Claremont, CA  91711; pic04747@pomona.edu}

\begin{abstract}

We present the results of a comprehensive Keck/DEIMOS spectroscopic
survey of the ultra-faint Milky Way satellite galaxy Segue~1.  We have
obtained velocity measurements for 98.2\% of the stars within 67~pc
(10\arcmin, or 2.3 half-light radii) of the center of Segue~1 that
have colors and magnitudes consistent with membership, down to a
magnitude limit of $r=21.7$.  Based on photometric, kinematic, and
metallicity information, we identify 71 stars as probable Segue~1
members, including some as far out as 87~pc.  After correcting for the
influence of binary stars using repeated velocity measurements, we
determine a velocity dispersion of $3.7^{+1.4}_{-1.1}$~\kms.  The mass
within the half-light radius is $5.8^{+8.2}_{-3.1} \times
10^{5}$~M$_{\odot}$.  The stellar kinematics of Segue~1 require very
high mass-to-light ratios unless the system is far from dynamical
equilibrium, even if the period distribution of unresolved binary
stars is skewed toward implausibly short periods.  With a total
luminosity less than that of a single bright red giant and a V-band
mass-to-light ratio of 3400~M$_{\odot}$/L$_{\odot}$, Segue~1 is the
darkest galaxy currently known.  We critically re-examine recent
claims that Segue~1 is a tidally disrupting star cluster and that
kinematic samples are contaminated by the Sagittarius stream.  The
extremely low metallicities ($\rm{[Fe/H]} < -3$) of two Segue~1 stars
and the large metallicity spread among the members demonstrate
conclusively that Segue~1 is a dwarf galaxy, and we find no evidence
in favor of tidal effects.  We also show that contamination by the
Sagittarius stream has been overestimated.  Segue~1 has the highest
estimated dark matter density of any known galaxy and will therefore
be a prime testing ground for dark matter physics and galaxy formation
on small scales.
\end{abstract}

\keywords{dark matter --- galaxies: dwarf --- galaxies: kinematics and
  dynamics --- galaxies: individual (Segue~1) --- Local Group}

\section{INTRODUCTION}
\label{introduction}
\setcounter{footnote}{8}

The Sloan Digital Sky Survey (SDSS) has been tremendously successful
in revealing new Milky Way dwarf galaxies over the past five years
\citep[e.g.,][]{willman05,zucker06,belokurov07,walsh07,belok10}.
However, its limited depth and sky coverage, along with the difficulty
of obtaining spectroscopic followup observations, still leave us with
an incomplete understanding of the Milky Way's satellite population.
In particular, key parameters such as the luminosity function, mass
function, radial distribution, and total number of satellites depend
extremely sensitively on the properties of the few least luminous
dwarfs \citep[e.g.,][]{tollerud08}, which are not yet well-determined.
Since the least luminous dwarfs are the closest and densest known dark
matter halos to the Milky Way, these same objects represent critical
targets for indirect dark matter detection experiments
\citep*[e.g.,][]{baltz00,evans04,colafrancesco07,strigari08b,kuhlen08,bringmann09,pieri09,martinez09}
and for placing limits on the phase space density of dark matter
particles
\citep[e.g.,][]{hd00,dh01,kaplinghat05,sg07,strigari08b,geha09}.
However, as the closest known satellites to the Milky Way, they are
also the most susceptible to tidal forces and other observational
systematics.

Because of the extreme lack of bright stars in these systems, most of
the faintest dwarfs such as Willman~1 \citep{willman05}, Bo{\"
  o}tes~II \citep{walsh07}, Segue~1 \citep{belokurov07}, and Segue~2
\citep{belokurov09} remain relatively poorly characterized by
observations; for example, the dynamical state of Willman~1 has still
has not been established \citep{martin07,willman10}, and the velocity
dispersion of Boo~II is uncertain at the factor of $\sim5$ level
\citep{koch09}.  Similarly, although \citet[][hereafter G09]{geha09}
demonstrated that the kinematics of stars in Segue~1 clearly indicate
that it is a dark matter-dominated object, other observations have
suggested the possibility of tidal debris in the vicinity of Segue~1,
as well as potential contamination from the Sagittarius stream
\citep{no09}.
 
More generally, the issues of tidal disruption \citep*[e.g.,][]{pmn08}
and binary stars \citep{mc10} are the last remaining major questions
to be settled regarding the nature of the faintest dwarfs.  These
objects promise clues to the extreme limits of galaxy formation
\citep{gilmore07,strigari08a} and perhaps to the formation of the
first galaxies in the early universe \citep[e.g.,][]{br09}, as well as
offering insights into dark matter physics.  However, these
applications hinge on the assumption that the mass distribution of
each system is accurately known.  Current mass estimates assume
dynamical equilibrium and that the observed kinematics are not being
affected by Galactic tides or binary stars, but tests of those
assumptions are obviously required in order to confirm that the dwarfs
are bound, equilibrium systems.  If instead the observed velocity
dispersions of Segue~1, Willman~1, and others are being inflated
either by the tidal influence of the Milky Way or the presence of
binary stars in the kinematic samples, then they are unlikely to be
useful probes of the behavior of dark matter on small scales.

Correcting velocity dispersions for binaries, which are inevitably
present in any stellar system, is relatively straightforward
\citep{minor10}.  The only observational requirement is that a
significant subset of the sample have at least two velocity
measurements with a separation of order 1~yr.  Tidal effects,
unfortunately, are more difficult to nail down.  The only unambiguous
signature of tidal interactions is the presence of tidal tails
\citep[e.g.,][]{toomre72}.  Detecting such features in the ultra-faint
dwarfs is extremely challenging: the galaxies themselves have central
surface brightnesses of $26-28$~mag~arcsec$^{-2}$ \citep*{martin08},
so any tidal debris would be at least several magnitudes fainter and
likely below the SDSS detection limit of $\sim30$~mag~arcsec$^{-2}$.
Deeper, wide-field photometric surveys of the ultra-faint dwarfs can
reach surface brightnesses as low as $32.5$~mag~arcsec$^{-2}$
\citep*{sand09,sand10,munoz10,dejong10}, but such observations are not
yet available for most of the dwarfs.

In principle, spectroscopic studies can pinpoint the stars associated
with an object and probe debris at lower surface densities than is
possible photometrically.  Spectroscopic surveys also provide the only
means of identifying tidal debris that is oriented along the line of
sight to an object \citep{lokas08,klimentowski09}.  However, the
currently available spectroscopic samples of less than 25 stars in the
faintest dwarfs are not sufficient to determine to what extent tides
may be affecting the kinematics.  Much larger spectroscopic data sets
are required to test for tidal effects.

In this paper, we present a nearly complete spectroscopic survey of
Segue~1 that is aimed at obtaining repeated velocity measurements of
known members and searching for stars that have been tidally stripped
from the system.  We describe our modeling of the binary star
population and the mass distribution in more detail in a companion
paper \citep[][hereafter Paper~II]{martinez}, and a separate study
examines the implications of our new mass measurements for indirect
detection of dark matter \citep{essig10}.  In Section~\ref{data} we
describe the survey and the data reduction.  We identify Segue~1
member stars in Section~\ref{member_sec}, and then analyze their
metallicities and velocities in Section~\ref{results}.  In
Section~\ref{sigma} we present our derivation of the intrinsic
velocity dispersion of Segue~1 after correcting for the presence of
binary stars in the sample (see Paper~II for more details), and in
Section~\ref{stream} we describe our detection of an unrelated tidal
stream in the same part of the sky.  We consider the implications of
this data set for proposals that the kinematics of Segue~1 are
affected by contamination and tidal disruption in Section~\ref{tides}.
We discuss the utility of Segue~1 for placing constraints on the
properties of dark matter in Section~\ref{darkmatter}.  In
Section~\ref{conclusions} we summarize our findings and conclude.

\section{EXPERIMENTAL DESIGN, OBSERVATIONS, AND DATA REDUCTION}
\label{data}

\subsection{A Survey for Tidal Debris}

As a complement to ongoing deep, wide-field photometric surveys of the
ultra-faint dwarfs \citep[e.g.,][]{munoz10}, we embarked upon a
spectroscopic search for evidence of tidal stripping or extratidal
stars.  The ideal target for such a search would be a galaxy that (1)
is nearby, to maximize the tidal forces it is currently
experiencing,\footnote{If the object is too close to the pericenter of
  its orbit, though, then the extent of its tails (if they exist)
  would be minimized.} (2) is moving at a high velocity relative to
the Milky Way, to minimize the degree of contamination by foreground
stars, and (3) has a small angular size, to minimize the area that the
survey needs to cover.  Out of all the known Milky Way dwarf galaxies,
the clear choice according to these criteria is Segue~1.  At a
distance of 23~kpc from the Sun (28~kpc from the Galactic Center),
Segue~1 is the closest dwarf galaxy other than Sagittarius, which of
course is the prototype for a dwarf undergoing tidal disruption.  Its
heliocentric velocity of 207~\kms\ (the largest of the Milky Way
satellites within 200~kpc) and relatively small velocity dispersion
give Segue~1 the lowest expected surface density of Milky Way
foreground stars within $3\sigma$ of its mean velocity (according to
the Besan{\c c}on model; \citealt{robin03}).  Finally, if Segue~1 is
\emph{not} surrounded by a massive dark matter halo --- and it can
only host visible tidal features if no extended halo is present ---
its instantaneous Jacobi (tidal) radius based on the stellar mass
estimated by \citet{martin08} is $\sim30$~pc, or 4.5\arcmin, which is
an observationally feasible area to search.  This calculation
conservatively assumes that Segue~1 has never been closer to the Milky
Way than it is now; if its orbital pericenter is less than 28~kpc, its
baryon-only tidal radius would be even smaller.  The properties of
Segue~1 are summarized in Table~\ref{segue1_table}.

\begin{deluxetable}{llr}
\tablecaption{Summary of Properties of Segue\,1}
\tablewidth{0pt}
\tablehead{
\colhead{Row} & \colhead{Quantity} & \colhead{Value}
}
\startdata
(1) & RA (J2000)             & 10:07:03.2$\pm1.7^{s}$ \\
(2) & Dec (J2000)      & +16:04:25$\pm15''$ \\
(3) & Distance (kpc)      & $23\pm2$  \\
(4) & $M_{V,0}$         & $-1.5^{+0.6}_{-0.8}$ \\
(5) & $L_{V,0}$ (L$_{\odot}$) & 340\\
(6) & $\epsilon$          & $0.48^{+0.10}_{-0.13}$ \\
(7) & $\mu_{V,0}$ (mag~arcsec$^{-2}$) &  $27.6^{+1.0}_{-0.7}$ \\
(8) & $r_{\rm eff}$ (pc)     & $29^{+8}_{-5}$  \\
\hline
(9)  & $V_{hel}$ (\kms)            & $208.5\pm0.9$\\
(10)  & $V_{\rm GSR}$ (\kms)       & $113.5\pm0.9$\\
(11)  & $\sigma$ (\kms)          & $3.7^{+1.4}_{-1.1}$\\
(12)  & Mass (M$_{\odot}$)       & $5.8^{+8.2}_{-3.1}\times10^5$  \\
(13)  & M/L$_{V}$ (M$_{\odot}$/L$_{\odot}$)       & 3400 \\
(14)  & Mean [Fe/H]       & $-2.5$
\enddata
\tablecomments{Rows (1)-(2) and (4)-(8) are taken from the SDSS
  photometric analysis of \citet{martin08} and row (3) from
  \citet{belokurov07}.  Values in rows (9)-(14) are derived in this
  paper.}
\label{segue1_table}
\end{deluxetable}

\subsection{Target Selection}
\label{targetselection}

\begin{figure}[t!]
\epsscale{1.24}
\plotone{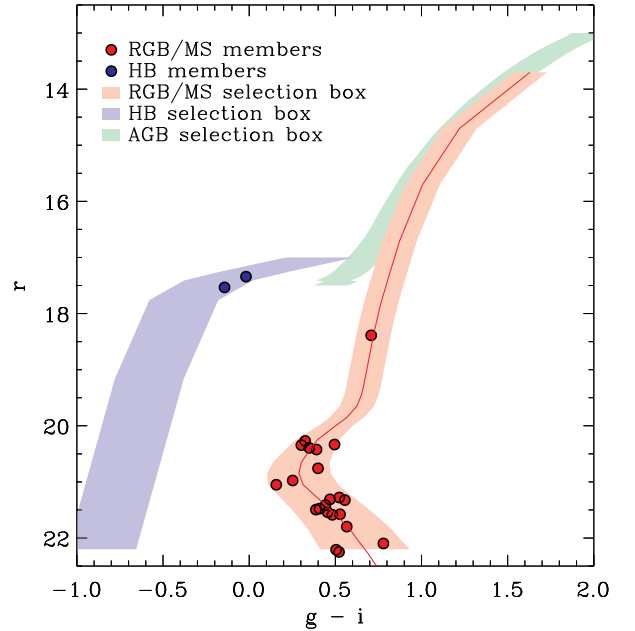}
\caption{Photometric selection criteria for candidate Segue~1 members.
  The red giant branch/main sequence selection region (shaded red) is
  based on the M92 isochrone of \citet{clem08}, adjusted slightly at
  magnitudes fainter than $r = 20.65$ so as to enclose all of the
  spectroscopically confirmed members from G09 (red line).  The blue
  and green shaded regions represent the horizontal branch (from the
  M13 isochrone of \citealt{clem08}) and asymptotic giant branch (from
  a \citealt{girardi04} theoretical isochrone at $\rm{[Fe/H]} = -1.7$)
  selection boxes, respectively.  The filled points are the 24 member
  stars identified by G09.}
\label{cmd_selection}
\end{figure}

To select targets for the survey, we focused on the area within
$\sim$15\arcmin\ (100~pc) of the center of Segue~1 as determined by
\citet{martin08}.  Guided by the 24 member stars identified by G09, we
tweaked the color of the (appropriately shifted and reddened) M92
isochrone from \citet{clem08} so that it passed through the center of
the member sequence at all magnitudes (this adjustment was only needed
for the subgiant branch and main sequence, not the red giant branch).
The Segue~1 main sequence appears to be slightly redder than that of
M92, with the offset increasing toward fainter magnitudes (see
Figure~\ref{cmd_selection}).  The full G09 member sample is located
within 0.25~mag of the adjusted fiducial track (0.2~mag for $r \le 21$
and 0.1~mag for $r \le 20$).

Using positions and magnitudes extracted from DR5 of the SDSS
\citep{sdss_dr5}, we selected stars within a narrow range of colors
around the adopted fiducial sequence.  Red giants ($r \le 20$) were
required to be within 0.1~mag of the sequence, and the selection
window was gradually widened toward fainter magnitudes, reaching
0.235~mag at $r = 21.7$.  Horizontal branch candidates were allowed to
be 0.2~mag away from the fiducial track.  A small number of stars
located near a metal-poor asymptotic giant branch isochrone from
\citet{girardi04} were also selected.  Within 10\arcmin\ (67~pc) of
Segue~1, we identified 112 stars lying within the color-magnitude
selection box (down to a magnitude limit of $r = 21.7$) that we
consider to be our primary target sample.  Stars up to a factor of two
farther away from the fiducial sequences or within the primary
color-magnitude selection region but at larger distances from Segue~1
were targeted with reduced priorities.  We also included as many of
the known member stars as possible on multiple slitmasks to obtain
repeated velocity measurements for constraining the binary population.

\subsection{Observations}
\label{observations}

We observed 12 new slitmasks, including at least one slit placed on
each of the 112 candidate member stars (plus repeat observations of 18
of the 24 members from G09), with the DEIMOS spectrograph
\citep{faber03} on the Keck~II telescope.  The observations took place
on the nights of 2009 February 18, 26, and 27 and 2010 February 12 and
13.  

The spectrograph setup and observing procedures were identical to
those described by SG07: we used the 1200~$\ell$/mm grating with an
OG550 filter to cover the wavelength range $6500-9000$~\AA\ at a
spectral resolution of $R=6000$ (slit width of 0.7\arcsec).  An
internal quartz lamp and Kr, Ar, Ne, and Xe arc lamps were employed
for flat-fielding and wavelength calibration, respectively.  Total
integration times for the science masks ranged from 10 minutes for a
mask targeting very bright stars to $\sim2$ hours for most of the
masks aimed at fainter stars.  The masks are summarized in
Table~\ref{masktable}.  Conditions during the observing nights were
generally good, with seeing ranging from $0.7 - 1.0$\arcsec\ and thin
cirrus at times.

\begin{deluxetable*}{lcccrcccc}
\tablecaption{Keck/DEIMOS Slitmasks}
\tablewidth{0pt}
\tablehead{
\colhead{Mask} &
\colhead{$\alpha$ (J2000)} &
\colhead{$\delta$ (J2000)} &
\colhead{PA} &
\colhead{$t_{\rm exp}$} &
\colhead{MJD of} &
\colhead{\# of slits} &
\colhead{\% useful} \\
\colhead{name}&
\colhead{(h$\,$ $\,$m$\,$ $\,$s)} &
\colhead{($^\circ\,$ $\,'\,$ $\,''$)} &
\colhead{(deg)} &
\colhead{(sec)} &
\colhead{observation} &
\colhead{} &
\colhead{spectra}
}
\startdata
Segue1-1\tablenotemark{a} & 10 07 06.01 & 16 02 56.1  & \phs\phn65.0 & 5400 & 54416.551 & 59 & 88\%\\
Segue1-2     & 10 07 08.85 & 16 04 51.9  & \phs180.0    & 7200 & 54881.311 & 65 & 83\%\\
Segue1-3     & 10 07 00.82 & 16 06 59.6  & \phn$-57.0$  & 5400 & 54881.416 & 61 & 92\%\\
Segue1-C     & 10 06 39.29 & 16 06 02.0  & \phs171.0    & 600  & 54881.284 & 38 & 87\%\\
segtide1     & 10 06 57.65 & 16 10 20.5  & \phs178.0    & 7500 & 54889.267 & 48 & 88\%\\
segtide2     & 10 06 36.07 & 16 07 39.6  & \phs171.0    & 7500 & 54889.343 & 48 & 90\%\\
segtide3     & 10 07 34.37 & 16 10 28.9  & \phs144.0    & 7500 & 54889.444 & 47 & 94\%\\
segtide4     & 10 06 19.07 & 16 01 34.7  & \phs179.0    & 7350 & 54890.243 & 50 & 84\%\\
segtide5     & 10 07 40.78 & 15 56 22.1  & \phs\phn\phn1.0 & 7800 & 54890.342 & 44 & 89\%\\
segtide6     & 10 06 59.81 & 15 56 50.8  & \phn$-78.0$  & 7200 & 54890.444 & 54 & 93\%\\
segtide7     & 10 07 14.20 & 15 55 43.9  & $-123.0$     & 2700 & 54889.542 & 49 & 76\%\\
segtide8     & 10 06 58.45 & 16 03 37.4  & $-134.0$     & 1200 & 55240.355 & 40 & 90\%\\
segtide9     & 10 07 07.83 & 15 56 24.7  & $-165.0$     & 1800 & 55240.402 & 48 & 85\%\\
\enddata
\tablenotetext{a}{Segue1-1 is the slit mask observed by G09.  For
  consistency, we use the velocities reported in that paper rather
  than re-reducing the mask and measuring the velocities again.}
\label{masktable}
\end{deluxetable*}

\subsection{Data Reduction}

As with the observations, data reduction followed the outline given in
SG07.  We used a modified version of the data reduction pipeline
developed for the DEEP2 galaxy redshift survey.  The additional
improvements we made to the code since SG07 were to re-determine the
wavelength solution for slits that were initially not fit well and to
identify and extract serendipitously observed sources more robustly.
The pipeline determines a wavelength solution for each slit,
flat-fields the data, models and subtracts the sky emission, removes
cosmic rays, coadds the individual frames, and then extracts the
spectra.

In the spring of 2009, the DEIMOS CCD array was experiencing an
intermittent problem wherein one of the eight chips (corresponding to
the blue half of the spectral range at one end of the field of view)
would fail to read out completely on some exposures.  In a few cases,
the affected chip was completely blank, while in others only the top
or bottom of the chip was absent and the remainder of the pixels were
saved normally.  We dealt with this problem conservatively by
excluding all of the data from the temperamental chip from our
reduction any time it exhibited abnormal behavior.  As a result, a
fraction of our targets have lower S/N in the blue than would be
expected from the total mask observing times, but since most of the
key spectral features for our analysis are on the red side of the
spectra, the overall impact is minor.

After reducing the data, we used the custom IDL code described by SG07
to measure the radial velocity of each star.  We first corrected each
spectrum for velocity offsets that could result from mis-centering of
the star in the slit by cross-correlating the telluric absorption
features against those of a telluric standard star specially obtained
for this purpose \citep[][SG07]{sohn07}.  The spectra were then
cross-correlated with a library of high S/N templates with well-known
velocities obtained with DEIMOS in 2006 and 2007.  The template
library contains 15 stars ranging from spectral type F through M,
mostly focusing on low-metallicity giants but also including
representative examples of subgiants, horizontal branch stars, main
sequence stars, and more metal-rich giants.  We also fit 4
extragalactic templates, identifying a total of 69 background galaxies
and quasars.  The velocity of each target spectrum is determined from
the cross-correlation with the best-fitting template spectrum.  We
estimate velocity errors using the Monte Carlo technique presented in
SG07: we add random noise to each spectrum 1000 times and then
redetermine its velocity in each iteration.  We take the standard
deviation of the Monte Carlo velocity distribution for each star as
its measurement uncertainty.  Previous analysis of a sample of stars
observed multiple times with the same instrument configuration and
analysis software indicates that our velocity accuracy is limited by
systematics at the 2.2~\kms\ level.  We have now confirmed the
magnitude of the systematic errors with a much larger sample of repeat
measurements than were used by SG07.

\subsection{Repeat Measurements}
\label{repeats}

\begin{figure*}[t!]
\epsscale{1.17}
\plottwo{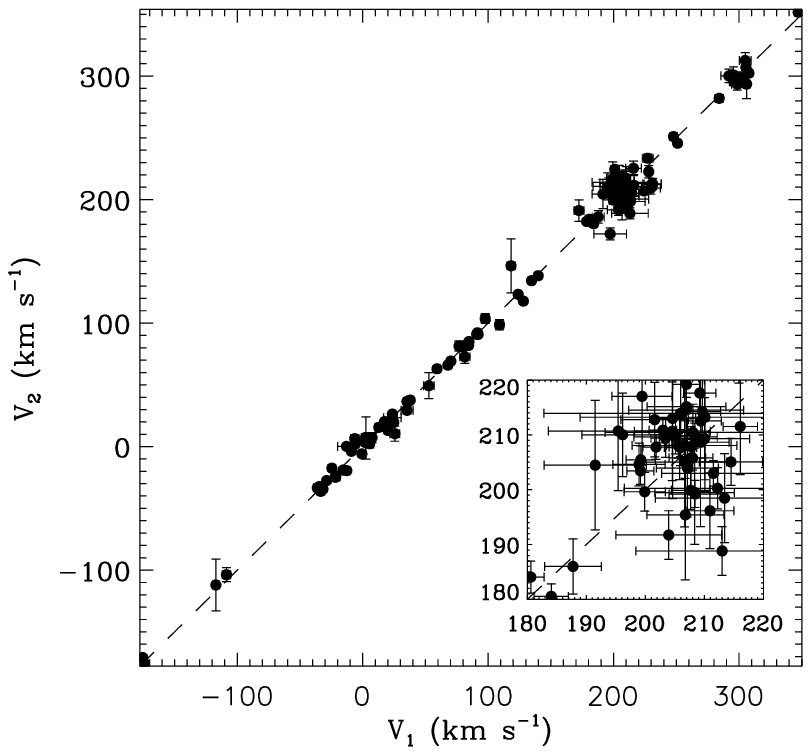}{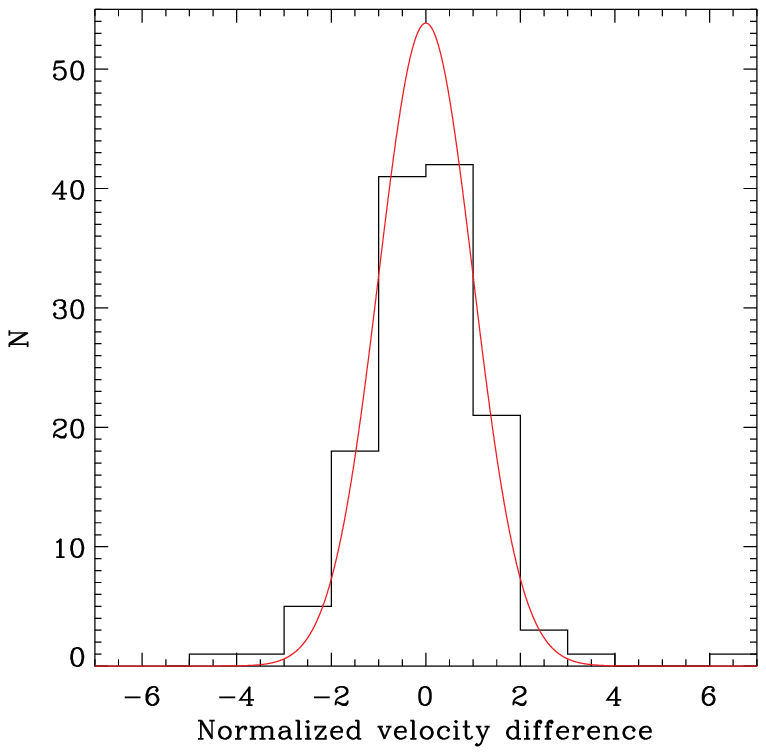}
\caption{(\emph{a}) Comparison of velocity measurements for the sample
  of stars observed at least twice.  The inset shows a zoomed-in view
  of the velocity range near Segue~1.  (\emph{b}) Histogram of
  velocity differences between repeat observations, normalized by
  their uncertainties ($(v_{2} - v_{1})/\sqrt{\sigma_{1}^2 +
    \sigma_{2}^2}$).  The red curve is a Gaussian with unit
  dispersion, which the data should follow if the measurement
  uncertainties are correct.  The larger than expected number of stars
  in the wings of the distribution is caused by the presence of
  binaries and RR~Lyrae variables in the sample.}
\label{repeat_fig}
\end{figure*}

As discussed in Sections~\ref{introduction} and \ref{targetselection},
one of the main goals of this study is to determine the effect of
binary stars on the observed velocity dispersion of Segue~1.
Accomplishing this task requires making multiple velocity measurements
of stars, verifying that the derived velocity uncertainties are
accurate, and searching for individual binaries.  The bulk of our
analysis of the repeat observations is presented in Paper II, but in
Figure~\ref{repeat_fig}\emph{a} we illustrate the agreement between
subsequent velocity measurements and the initial one for each star
observed more than once.  Figure~\ref{repeat_fig}\emph{b} shows the
distribution of velocity differences for each pair of measurements,
normalized by their uncertainties.  The good match to a Gaussian
distribution for $-2 < (v_{2} - v_{1})/\sqrt{\sigma_{1}^2 +
  \sigma_{2}^2} < 2$ indicates that the uncertainties are accurate,
and the excess in the wings of the distribution provides evidence for
velocity variability (see \S~\ref{binaries} and Paper~II).  A total of 93
stars, including approximately half of the member sample, were
observed at least twice during the course of our survey.

\subsection{Spectroscopic Completeness}
\label{completeness}

In \S\,\ref{observations}, we described obtaining a spectrum of each
one of the 112 stars within our primary photometric selection region
and not more than 67~pc from the center of Segue~1.  We successfully
measured velocities for 109 of these stars.  One target star
(SDSSJ100707.12+160022.4) did not have any identifiable features in
its spectrum, one spectrum (SDSSJ100700.75+160300.5) suffered from
reduction difficulties (see Appendix), although it appears to be a member,
and the remaining one (SDSSJ100733.12+155736.7) was not
detected in our data.  The magnitude of this latter star should have
made it easily visible in the exposures we obtained, but a check of
the SDSS images showed no source at this position, and the target was
flagged in the SDSS catalog as a possible moving object.  We conclude
that this source was actually an asteroid that happened to have the
right colors and position to be selected for our survey.  After
excluding it, our completeness is 98.2\%.  We also targeted a fraction
of Segue~1 member candidates as far out as 16\arcmin\ (107~pc).
Within 11\arcmin, 12\arcmin, and 13\arcmin, our overall completeness
is 97.0\%, 96.7\%, and 92.3\%, respectively.  Only a few stars at
larger radii were observed.

\section{DEFINING THE SEGUE~1 MEMBER SAMPLE}
\label{member_sec}

In total (including the observations of G09) we obtained 522 good
spectra of 393 individual stars, of which 167 were classified as
high-priority member candidates (109 within 10\arcmin\ of the center
of Segue~1 and 58 at larger radii) according to the criteria described
in \S\,\ref{targetselection}.  We present all of our velocity and Ca
triplet equivalent width measurements in Table~3.  Our repeat velocity
measurements for 93 of these stars span a maximum time baseline of
2.25 years.  We use a variety of techniques to identify Segue~1 member
stars in this data set.  The primary data available for distinguishing
members from non-members are: color/magnitude, velocity, metallicity
(either in the form of [Fe/H] estimated from the Ca triplet lines or
simply the raw Ca triplet equivalent width), spatial position, and the
strength of the \ion{Na}{1} $\lambda$8190~\AA\ doublet.\footnote{We
  note, however, that the Na lines only function as a way to
  distinguish dwarfs from giants for very red stars ($V-I \gtrsim 2$),
  significantly redder than anything we expect to find in a
  low-luminosity, low-metallicity system like Segue~1
  \citep{gilbert06}.  Also, because of the tiny number of giants in
  Segue~1, the majority of the \emph{member} stars are actually on the
  main sequence rather than the giant branch.  Therefore, the strength
  of the Na doublet primarily serves to eliminate stars that were
  already obvious non-members from consideration.}

\subsection{Methods of Identifying Segue~1 Members}
\label{selection}

We consider three different membership selection techniques: a
subjective, star-by-star selection using velocity, metallicity, color,
magnitude, and the spectrum itself, a slightly modified version of the
algorithm introduced by \citet{walker09}, and a new Bayesian approach
presented in Paper~II.  For the remainder of the paper, we adopt the
third method as defining our primary sample except where explicitly
noted.

\begin{figure*}[th!]
\epsscale{1.20}
\plotone{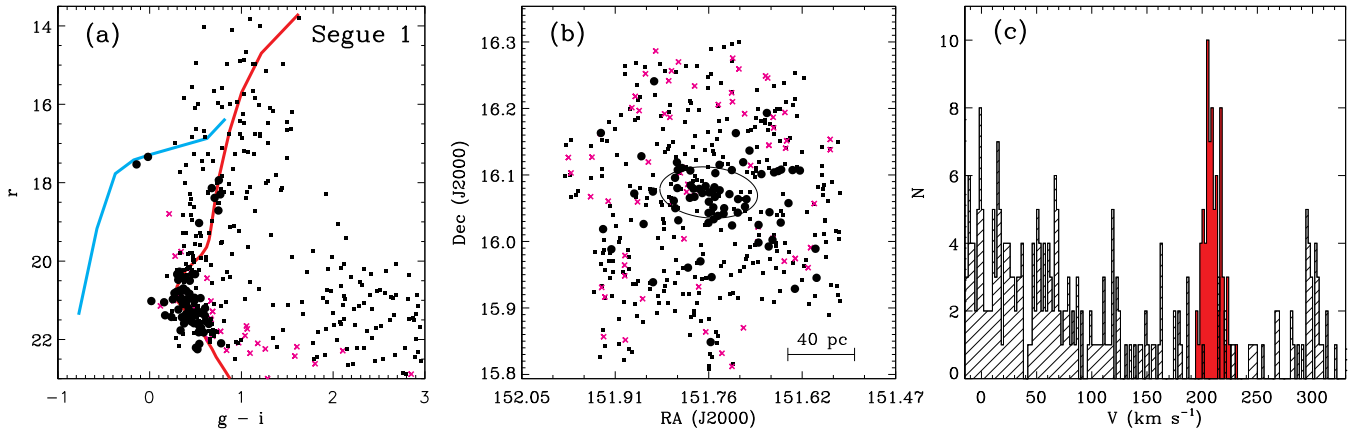}
\caption{(\emph{a}) Color-magnitude diagram of observed stars in
  Segue~1.  The large black circles represent the 71 stars identified
  as definite or probable radial velocity members of the galaxy using
  our subjective approach, the small black dots represent stars
  identified as likely or certain non-members, and the magenta crosses
  are spectroscopically confirmed background galaxies and quasars.
  The red curve shows the location of the red giant branch, subgiant
  branch, and main sequence turnoff populations in the globular
  cluster M92 and the cyan curve shows the location of the horizontal
  branch of M13, both corrected for Galactic extinction and shifted to
  a distance of 23 kpc \citep[data from][]{clem08}.  (\emph{b})
  Spatial distribution of observed stars in Segue~1.  Symbols are the
  same as in (\emph{a}), and the ellipse represents the half-light
  radius of Segue~1 from \citet{martin08}.  (\emph{c}) Velocity
  histogram of observed stars in Segue~1.  Velocities are corrected to
  the heliocentric rest frame.  The filled red histogram represents
  stars classified as members, and the hatched black-and-white
  histogram represents non-members.  The velocity bins are
  2~\kms\ wide. }
\label{cmd_radec_vhist_fig}
\end{figure*}

We first select member stars using the parameters listed above.
Candidate members must have colors and magnitudes consistent with the
photometric selection region displayed in Figure~\ref{cmd_selection}
and described in \S\,\ref{targetselection}, must have velocities
within a generous $\sim4\sigma$ window around Segue~1 based on the
systemic velocity and velocity dispersion measured by G09, and should
have low metallicities.  We can then iteratively refine the member
selection by examining the stars near the boundaries of the selection
region more carefully.  With this process, we classify 65 stars as
definite members, 6 additional stars as probable members, and 5 more
as likely (but not certain) non-members.  The remaining stars are
clearly not members.  The sample selected in this way is displayed in
Figure~\ref{cmd_radec_vhist_fig}.

While the flexibility afforded by this subjective approach is useful
and allows all available information to be taken into account, a
rigorous and objective method is also desirable to avoid the
possibility of bias.  \citet{walker09} recently developed such a
statistical algorithm to separate two potentially overlapping
populations.  This technique, known as expectation maximization (EM),
allows one to estimate the parameters of a distribution in the
presence of contamination, with specific application to the case of
identifying dwarf galaxy member stars against a Milky Way foreground.
Given a set of velocities, radial distances, and a third parameter in
which dwarf galaxy stars and Milky Way stars are distributed
differently, the EM algorithm relies on the distinct distributions of
the two populations in each property to iteratively assign membership
probabilities to each star until it converges on a solution.
\citet{walker09} use the pseudo-equivalent width of the Mg triplet
lines at 5180~\AA\ as their third parameter, but we find that the
reduced equivalent width of the Ca triplet (CaT) lines works as well.
\citet{rutledge97a,rutledge97b} define this reduced equivalent width
as $W' = \Sigma \rm{Ca} - 0.64(V_{\rm{HB}} - V)$, where $\Sigma
\rm{Ca}$ is the weighted sum of the equivalent widths (EWs) of the CaT
lines: $\Sigma \rm{Ca} = 0.5\rm{EW_{8498}} + 1.0\rm{EW_{8542}} +
0.6\rm{EW_{8662}}$.  Note that since the EM algorithm does not
incorporate photometric information, it must be run on a sample of
stars that has already passed the color-magnitude selection.  Also, EM
may fail to select horizontal branch (HB) stars because their broad
hydrogen lines interfere with measurements of the CaT EW; fortunately,
the HB members of Segue~1 are obvious.

\begin{figure*}[t!]
\epsscale{1.20}
\plotone{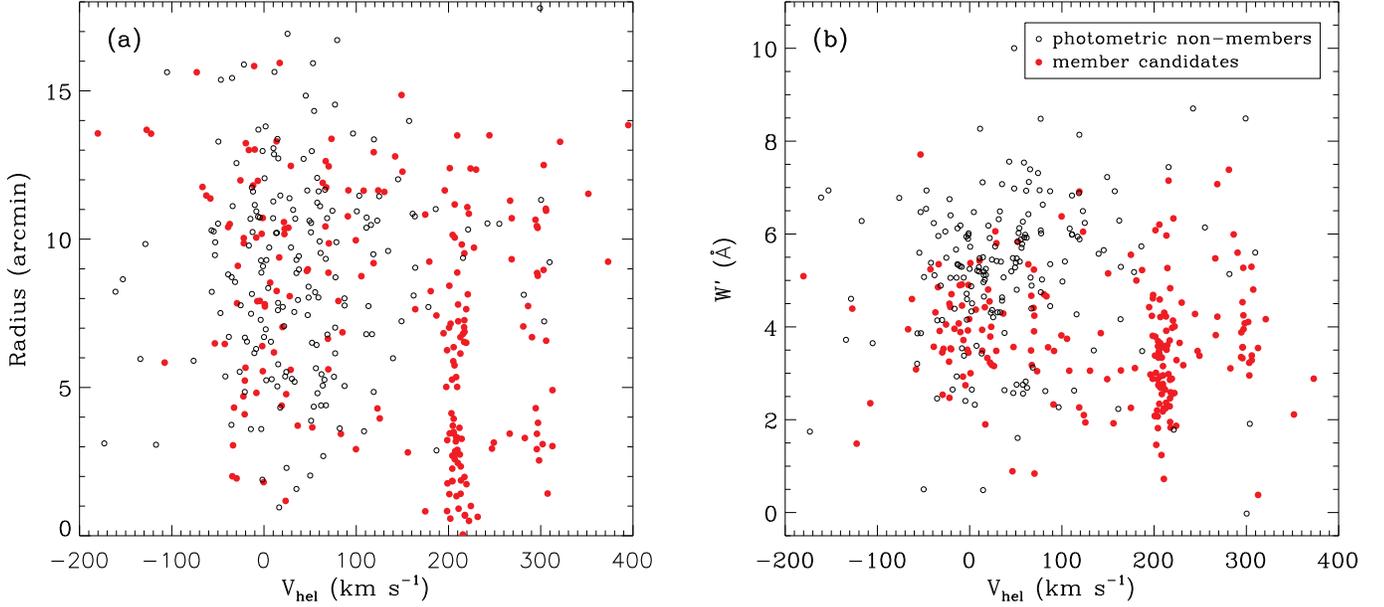}
\caption{(\emph{a}) Distribution of observed stars in velocity and
  radius.  Filled red points represent stars that pass the color and
  magnitude selection (at either high or low priority) described in
  \S\,\ref{targetselection}, and open black points are stars that lie
  outside that selection region.  Stars that have been observed
  multiple times are plotted with their weighted average values.
  Segue~1 stands out as the large overdensity of stars near $v_{hel} =
  200$~\kms\ extending out to a radius of $\sim13$\arcmin.  Based on
  the distribution of Milky Way stars, it is clear that at small radii
  ($r \le 7$\arcmin) the level of contamination of the Segue~1 member
  sample is very low.  In addition to Segue~1, there is also a
  distinct concentration of stars near 300~\kms.  (\emph{b})
  Distribution of observed stars in velocity and reduced Ca triplet
  equivalent width, a proxy for metallicity.  As in the left panel, a
  large fraction of the Segue~1 members separate cleanly from the
  Milky Way foreground population.  At $\rm{W'} > 5$~\AA, the
  distributions begin to overlap, and unambiguously classifying
  individual stars as members or nonmembers becomes more difficult.
  Fortunately, relatively few stars are located in this region.  It is
  clear that Segue~1 is more metal-poor than the bulk of the
  foreground population, although $\rm{W'}$ is a much less accurate
  metallicity indicator for main sequence stars than giants.  The
  300~\kms\ structure appears to be more enriched than Segue~1.  }
\label{r_v_ew}
\end{figure*}

\begin{figure*}[t!]
\epsscale{1.20}
\plotone{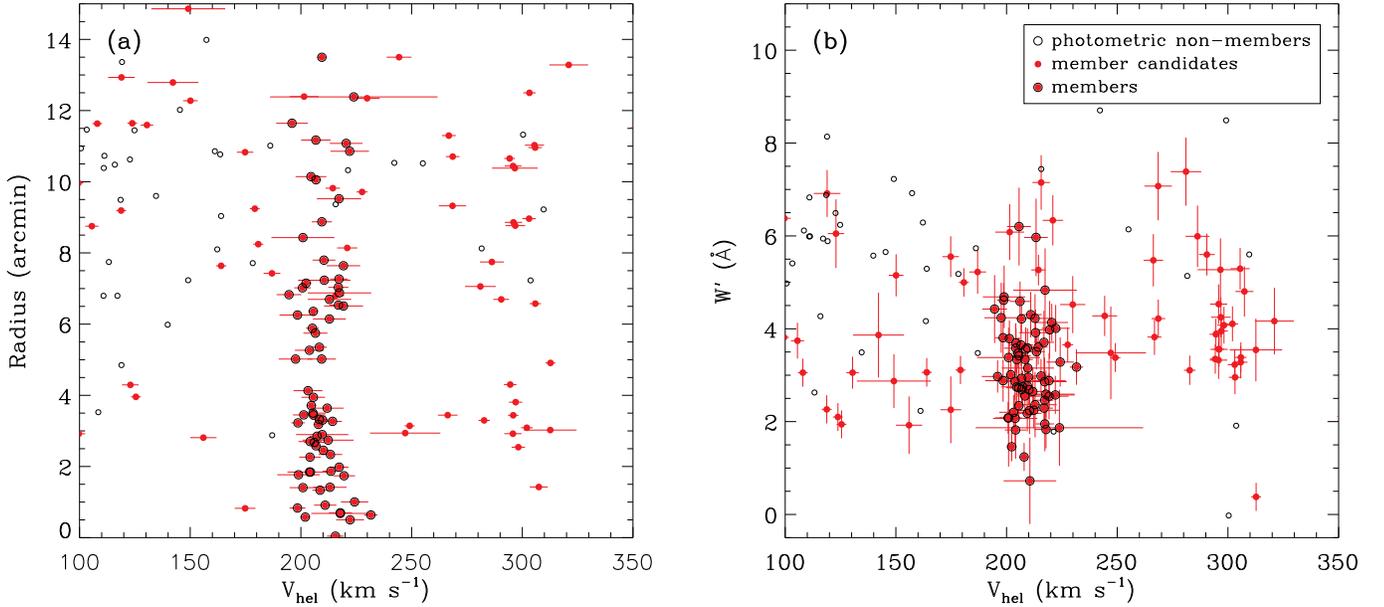}
\caption{(\emph{a}) Distribution of observed stars in velocity and
  radius, zoomed in on Segue~1 and the 300~\kms\ stream.  Symbols are
  as in Figure~\ref{r_v_ew}, but we have added error bars in velocity
  and highlighted the subjective 71 star Segue~1 member sample (filled
  red circles outlined in black).  (\emph{b}) Distribution of observed
  stars in velocity and reduced Ca triplet equivalent width, zoomed in
  on Segue~1 and the 300~\kms\ stream. }
\label{r_v_ew_mem}
\end{figure*}

We display the distribution of observed stars in radius, velocity, and
reduced CaT EW in Figures \ref{r_v_ew} and \ref{r_v_ew_mem}.  Segue~1
stands out as the large overdensity of stars with velocities just
above 200~\kms\ and smaller than average radii and $W'$ values.  We
caution that $W'$ is only a properly calibrated metallicity indicator
for stars on the red giant branch (RGB), which constitute a small
minority of the data set examined here.  Nevertheless, experiments
with globular cluster stars reaching several magnitudes below the main
sequence turnoff from the compilation of \citet{kirby10} show that
while $W'$ does increase at constant metallicity toward fainter main
sequence magnitudes, this increase is less than 2~\AA\ for stars
within 2 magnitudes of the turnoff.  We therefore conclude that
including both RGB and main sequence stars may broaden the Segue~1
distribution towards higher values of $W'$ (perhaps accounting for the
clear presence of Segue~1 stars in Figure~\ref{r_v_ew}b up to $W'
\approx 6$~\AA), but should not significantly affect the performance
of EM.

The EM algorithm selects 68 stars as definite members of Segue~1
(membership probability $p \ge 0.9$).  An additional 3 stars have $0.8
\le p < 0.9$ and are classified as members by eye, yielding 71 very
likely members.  These 71 stars correspond to 70 of the 71
subjectively classified members.\footnote{The EM algorithm includes
  one star (SDSSJ100711.80+160630.4) with a velocity of $247.1 \pm
  15.9$~\kms\ that we rated as too far removed from the systemic
  velocity to be a member candidate, and gives one star
  (SDSSJ100622.85+155643.0) with a closer velocity but an even larger
  uncertainty ($v_{hel} = 223.9 \pm 37.8$~\kms) a lower membership
  probability of $p = 0.73$.  Both of these stars are discussed
  further in the Appendix.}  Finally, 3 stars have membership
probabilities of $0.5 \le p < 0.8$.

Our final results are based on a Bayesian analysis that allows for
both contamination by Milky Way foreground stars and the contribution
of binary orbital motions to the measured velocities.  These
calculations are a natural generalization of the \citet{walker09} EM
method.  The method is described in more detail in Paper~II and is
summarized here in \S\,\ref{sigma}.  In this framework, we find 53
definite members ($\langle p\rangle \ge 0.9$) and 9 further probable
members ($0.8 \le \langle p\rangle < 0.9$), plus the 2 RR~Lyrae
variables (see \S\,\ref{binaries}), but 7 of the stars considered
likely members by the other two techniques receive lower probabilities
of $0.4 \le \langle p\rangle < 0.8$ here.  With the exception of the
discussion in \S\,\ref{kinematics}, where we mention the range of
velocity dispersions that can be obtained for different member
samples, the main results of this paper (including the velocity
dispersion, mass, and density of Segue~1) rely on this Bayesian
analysis.  It is important to note that unlike previous studies,
\emph{we include all stars that pass our photometric cuts in the
  Bayesian calculations, not just the ones with high membership
  probabilities.}  Each star is weighted according to its probability
of being a member of Segue~1.  This approach allows us to account
correctly for the significant number of stars with membership
probabilities that are neither close to zero nor close to one.

\section{METALLICITY AND KINEMATICS OF SEGUE~1}
\label{results}

\subsection{Stellar Metallicities and the Nature of Segue~1}

One of the defining differences between galaxies and globular clusters
is that dwarf galaxies universally exhibit signs of internal chemical
evolution and contain stars with a range of metallicities, while
globulars generally do not.  Recent work has shown that multiple
stellar populations with different chemical abundance patterns are in
fact present in some globular clusters, but these differences tend to
be subtle (which is why they are only being recognized now) and are
preferentially found in luminous clusters that are often argued to be
the remnants of tidally stripped dwarf galaxies
\citep[e.g.,][]{lee99,marino09,dacosta09,ferraro09,cohen10}.
\label{metallicity}

We use the spectral synthesis method introduced by \citet*{kgs08} and
refined by \citet{kirby09,kirby10} to measure iron
abundances\footnote{For historical reasons, these calculations use a
  solar iron abundance of $12 + \log{\epsilon{\rm(Fe)}} = 7.52$
  \citep{ag89}, but the difference between this assumption and the
  modern value of $12 + \log{\epsilon{\rm(Fe)}} = 7.50 \pm 0.04$
  \citep{asplund09} is negligible.} in Segue~1 directly from our
medium-resolution spectra.  \citet{kirby10} showed that the
metallicities measured in this way are reliable for stars with
$\log{g} < 3.6$.  Thus, we can only determine metallicities for the 6
red giant members of Segue~1; the fainter stars are all at or below
the main sequence turnoff.  The metallicities of these 6 stars span an
enormous range, with two stars at $\rm{[Fe/H]} > -1.8$ and two others
at $\rm{[Fe/H]} < -3.3$ (see Table~\ref{metallicity_table}).  One of
the two extremely metal-poor (EMP) stars does not have a well-defined
metallicity measurement because no Fe lines are detected in its
spectrum.  The upper limit on its metallicity therefore depends on the
assumptions, but it is certainly well below $\rm{[Fe/H]} = -3$.  The
mean metallicity of the Segue~1 red giants is $\rm{[Fe/H]} = -2.5$,
comparable to the most metal-poor galaxies identified so far
\citep{kirby08}, and the standard deviation, while not
well-constrained with such a small sample, is $\sim0.8$~dex.  Using a
completely independent data set, \citet{norris10b} reach essentially
identical conclusions regarding the Segue~1 abundance range and
identify yet another EMP member star at $\rm{[Fe/H]} = -3.5$
\citep{norris10a}.

The very large star-to-star spread in metallicities and the presence
of EMP stars with $\rm{[Fe/H]} < -3.0$ each independently argue that
Segue~1 cannot be a globular cluster, contrary to initial suggestions
\citep{belokurov07,no09}.  Only $\omega$~Centauri among globular
clusters has a comparable metallicity spread
\citep[e.g.,][]{ndc95,hilker04}, and that object is widely regarded to
be the remnant of a dwarf galaxy
\citep*{lee99,majewski00b,cl00,hr00,tsuchiya03,mizutani03,rey04,im04,msh05,carretta10}
rather than a true globular.  The lowest metallicity Segue~1 giants
are also at least a factor of four more metal-poor than any known star
in a globular cluster
\citep[e.g.,][]{king98,ki03,preston06,carretta09}.  We conclude that
the metallicities of stars in Segue~1 provide compelling evidence
that, irrespective of its current dynamical state, Segue~1 was once a
dwarf galaxy.

\subsection{Binary and Variable Stars}
\label{binaries}

Before attempting to determine the velocity dispersion and mass of
Segue~1, we consider the impact of variable stars and binaries that
could be in our sample.  The G09 members include two horizontal branch
stars in Segue~1.  Our repeated measurements demonstrate that the
velocities of both of these stars vary with time, leading us to
conclude that they are RR~Lyrae variables.  Followup photometry with
the Pomona College 1~m telescope at Table Mountain Observatory
confirms that one of these stars, SDSSJ100644.58+155953.9, is a
photometric variable with a characteristic RR~Lyrae period of
0.50~days.\footnote{Periods of 0.5~d are of course subject to the
  possibility of aliasing, and the light curve is not complete enough
  to rule out a period of 1.0~d.  However, such long periods are
  extremely rare for RR~Lyraes, so we consider the 0.50~d period to be
  the most likely solution.} We do not detect variability in the
second star, SDSSJ100705.60+160422.0, but the limits we can place are
not inconsistent with the low amplitude variability that might be
expected for such a blue star.  Given their blue colors, the stars are
probably type c variables pulsating in the first overtone mode.
Because the light curve phases at the times our spectra were acquired
are not known, we cannot measure the center-of-mass velocities of
these stars and must remove them from our kinematic sample even though
they are certainly members of Segue~1.

We also obtained multiple measurements of five of the six Segue~1 red
giants in order to check whether any of them are in binary systems.
For four of the stars, the velocity measurements do not deviate by
more than $2~\sigma$ from each other, providing no significant
indication of binarity, although long period or low amplitude orbits
cannot be ruled out.  SDSSJ100652.33+160235.8, however, shows clear
radial velocity variability, with the velocity decreasing from $216.1
\pm 2.9$~\kms\ on 2007 November 12 to $203.0 \pm 2.3$~\kms\ on 2009
February 27, and then rising back to $210.8 \pm 2.3$~\kms\ on 2010
February 13.  Assuming that these observations correspond to a single
orbital cycle, we infer a period of $\sim1$~yr and a companion mass of
$\sim 0.65 + 0.25(1/\mbox{sin}^{3}i - 1)$~M$_{\odot}$.

With at least one out of the brightest six stars (excluding the even
more evolved horizontal branch stars) in the galaxy in a binary
system, the binary fraction of Segue~1 is likely to be significant, as
has been found for other dwarf galaxies \citep*{queloz95,olszewski96}
and some, although not all, globular clusters
\citep*{fischer93,yc96,rb97,clark04,sollima07}.  For main sequence
stars, which dominate our sample, the binary fraction can only be
larger since the tight binary systems will not yet have been destroyed
by the evolution of the more massive component.  We have a limited
sample of repeat observations of some of the main sequence members, in
which two additional stars, SDSSJ100716.26+160340.3 and
SDSSJ100703.15+160335.0, are detected as probable binaries.  However,
these binary determinations are almost certainly quite incomplete, and
proper corrections for the inflation of the observed velocity
dispersion of Segue~1 by binaries must be done in a statistical sense,
as we discuss in \S\,\ref{sigma}.

\subsection{Kinematics}
\label{kinematics}

Because the issues of membership and binary stars are so critical to
our results, we must experiment with different samples of member stars
and methods of determining the velocity dispersion.  Inspection of
Figure~\ref{r_v_ew} makes clear that for $W' \le 3$~\AA, the expected
contamination by Milky Way foreground stars is negligible ($1-2$
stars).  Very conservatively, then, we can select the stars with $W'
\le 3$~\AA\ and 190~\kms~$ \lesssim v \lesssim $~225~\kms\ as an
essentially clean member sample (the exact velocity limits chosen do
not matter, since the next closest stars are at $v = 175$~\kms\ and $v
> 300$~\kms).  With the two RR Lyrae variables and the one obvious RGB
binary removed, the velocity dispersion of the other 34 stars is $3.3
\pm 1.2$~\kms.  (Note that we calculate the velocity dispersion using
a maximum likelihood method following \citealt{walker06}.)  This value
can be regarded in some sense as a lower limit to the observed
dispersion of Segue~1 (prior to any correction for undetected
binaries).

Since this conservative approach involves discarding nearly half of
the data, we would also like to consider alternatives.  The largest
member sample that we can define is the 71 stars selected using either
our holistic, subjective criteria in \S\,\ref{selection} or the EM
algorithm.  After again excluding the two RR Lyraes and the RGB
binary, the raw velocity dispersion of these stars is $5.5 \pm
0.8$~\kms, which we take as an \emph{upper} limit to the observed
dispersion of Segue~1 (as before, prior to correcting for undetected
binaries).  However, the dispersion in this case is dominated by a
single star (SDSSJ100704.35+160459.4; see \S~\ref{ambiguous}).  If we
remove this object from the sample, the dispersion of the remaining
stars falls to $3.9 \pm 0.8$~\kms, corresponding to a factor of 2
decrease in the mass of the galaxy.  Other stars that are borderline
members or non-members have negligible effects on the derived
dispersion (see Appendix).

We note at this point a curious finding regarding the brightest stars
in Segue~1.  If we isolate the evolved stars (6 giants and 2
horizontal branch stars) in the sample, their velocity dispersion
\emph{appears} to be quite small.  For the HB stars and the binary on
the giant branch we cannot assume that we have enough measurements to
average out the effects of the binary orbit and RR~Lyrae pulsations,
but we estimate a dispersion of $1.3^{+2.4}_{-0.7}$~\kms\ for the
other five RGB stars.  Using our full Bayesian analysis
(\S~\ref{sigma}) and including the binary, the intrinsic dispersion of
the giants is $2.0^{+3.1}_{-1.7}$~\kms.  Given the substantial error
bars, these values are formally consistent with the larger dispersion
obtained for the full data set, even though they are also close to
zero.  Nevertheless, the velocity dispersion we determine for the
remaining stars is not significantly affected by the inclusion or
exclusion of the giants and horizontal branch stars.  Without any
known physical mechanism that could change the kinematics of Segue~1
for stars in different evolutionary states, we conclude that the
apparently small dispersion of these stars is most likely a
coincidence resulting from small number statistics.

Next, we use the sample defined by the \citet{walker09} EM algorithm.
Since this sample is nearly identical to that considered in the
previous paragraphs, the results are unchanged: a dispersion of $5.7
\pm 0.8$~\kms\ for all 71 stars minus the RR Lyraes and the RGB
binary, and $4.1 \pm 0.9$~\kms\ when SDSSJ100704.35+160459.4 is
removed.  It is worth noting that all of these measurements and those
described above are consistent within $1\sigma$ with the original
velocity dispersion of $4.3 \pm 1.2$~\kms\ determined by G09.

The two above methods make various assumptions about how membership is
defined, but do not allow for a fully self-consistent statistical
treatment.  The method described in Paper~II \citep{martinez} and
summarized in \S\,\ref{sigma} treats these assumptions and the data
analysis in a fully Bayesian manner.  This analysis identifies a total
of 61 stars (excluding the 2 RR Lyrae variables) as likely members
with $\langle p\rangle > 0.8$.  This method arrives at significantly
lower membership probabilities for 8 stars compared to the EM and
subjective analyses.  These stars fall into three partially
overlapping categories: velocity outliers, frequently with large
velocity uncertainties as well (such that there is a significant
chance that the star's true velocity is far away from the systemic
velocity of Segue~1); stars with large reduced CaT EWs ($W' > 4$~\AA);
and stars at large radii ($r > 10$\arcmin).  The only one of these
stars that has an appreciable effect on the velocity dispersion is
SDSSJ100704.35+160459.4.  Removing each of the 7 other stars from the
sample changes the dispersion by less than 0.2~\kms.  With only the 61
most likely members included in the calculation, we find a velocity
dispersion of $3.4 \pm 0.9$~\kms.

\subsection{Individual Stars With Ambiguous Membership}
\label{ambiguous}

Despite our best efforts to define a member sample that is both clean
and complete, there are fundamental uncertainties that cannot be
avoided in determining whether any given star is a member of Segue~1.
In particular, we know that Segue~1 is an old, metal-poor stellar
system located 23~kpc from the Sun, and moving at a heliocentric
velocity of $\sim207$~\kms.  Unfortunately, the Milky Way halo also
contains old, metal-poor stars that span ranges in distance and
velocity that encompass Segue~1.  Given a large enough search volume,
it is therefore inevitable that some halo stars with the same age,
metallicity, distance (and hence the same colors and magnitudes), and
velocity as Segue~1 will be found.  We can use observations and models
to estimate the \emph{expected number} of such stars included in our
survey, but that does not help us in ascertaining the provenance of
individual stars.

As a result, we are left with a small number of stars whose membership
status is necessarily uncertain.  The algorithms discussed in
\S\,\ref{selection} allow us to assign membership probabilities to
these objects, which is the best statistical way to deal with our
limited knowledge.  Nevertheless, each star that we observed either is
or is not a member, and with a small sample, the assumption that a
particular star has a membership probability of, e.g., 0.6 can produce
different results than if it were known absolutely to be a member or
not.  Most of the stars in this category do not have an appreciable
effect on the derived properties of Segue~1 (most importantly the
velocity dispersion), either because they lie near the middle of the
distribution or because their velocity uncertainties are relatively
large.  One object, however, can make a significant difference, as
discussed in the following paragraph.  A few other stars that cannot
be classified very confidently in one category or the other are listed
in the Appendix, but their inclusion or exclusion has minimal impact
on the properties of Segue~1 or the results of this paper.

The one star that can individually affect the kinematics of Segue~1 is
SDSSJ100704.35+160459.4.  This star is a major outlier in velocity,
with a mean velocity from two measurements of $231.6 \pm 3.0$~\kms.
It is therefore located more than $6\sigma$ (where $\sigma$ is
estimated from the rest of the stars) away from the systemic velocity
of Segue~1, but because its position is extremely close (38\arcsec) to
the center of the galaxy and it has a CaT EW that could plausibly
be associated with Segue~1 (although on the high side), the EM method
returns a membership probability of 1.  The membership probability
from our full Bayesian analysis (see Paper~II) is lower, but far from
negligible, at 0.49.  The high velocity of this star relative to the
systemic velocity of Segue~1 could be explained if it is a member of a
binary system, but our two velocity measurements of it (separated by
1~yr) do not show a significant change in velocity, so the period
would have to be $\gtrsim5$~yr.  In addition to its disproportionate
effect on the velocity dispersion, for an equilibrium model a star
that is a $6\sigma$ outlier from the mean velocity but is located so
close to the center of the galaxy implies strongly radial orbits.
None of the other stars in the sample lead to a preference for extreme
velocity anisotropy.

\section{The Intrinsic Velocity Dispersion and Mass of Segue~1}
\label{sigma}

Having carefully considered the membership of each star and the
effects of the key outliers in the previous two sections, we are now
in position to determine the best estimate of the intrinsic velocity
dispersion of Segue~1 based on all available data.  The full
calculations that we use for this purpose are presented in
\citet{martinez}, but we include a summary here for convenience.

\subsection{Binary Correction Method}
\label{binary_methods}

Given a sample of stars that may be members of Segue~1, we allow for
the possibility that some of the observed stars are likely members of
binary star systems rather than single stars.  We therefore must treat
the (unknown) velocity of the star system's center of mass $v_{cm}$
and the measured velocities themselves as distinct quantities.  For
each star system of absolute magnitude $M_{V}$ we have a set of
measured velocities $v_{i}$, where $i$ runs over the number of repeat
observations of the star under consideration at times $t_{i}$, and the
associated measurement uncertainties $e_{i}$.  We write the likelihood
of obtaining the observed data for each star assuming it is a member
of Segue 1 (S1) in terms of a joint probability distribution in the
measured velocities $v_i$ and the unknown center-of-mass velocity
$v_{cm}$:

\begin{eqnarray}
\lefteqn{\mathcal{L}_{\mathrm{S1}}(v_{i}|e_{i},t_{i},M_{V};\sigma,\mu,B,\mathscr{P})}
\nonumber \\ & = & \int_{-\infty}^{\infty}
P(v_{i}|v_{cm},e_{i},t_{i},M_{V};B,\mathscr{P})P(v_{cm}|\sigma,\mu)
dv_{cm} \label{overall_likelihood} \\ 
& \propto & (1-B) \frac{e^{-\frac{(\langle
      v\rangle-\mu)^{2}}{2\sigma^{2}}}}{\sqrt{\sigma^2 +
    \sigma_{m}^2}} + BJ(\sigma,\mu|\mathscr{P}), \nonumber
\end{eqnarray}

\noindent
where the first factor in the integrand is the probability of drawing
a set of velocities $v_{i}$ given a center-of-mass velocity $v_{cm}$
and certain values for the binary parameters $B$ and $\mathscr{P}$.
$B$ represents the binary fraction and $\mathscr{P}$ is the set of
binary parameters $[\mu_{\log P},\sigma_{\log P}]$ (see below).  The
second factor is the probability distribution of center-of-mass
velocities given an intrinsic velocity dispersion of Segue~1,
$\sigma$, and a systemic velocity, $\mu$.  In the last line of
Equation~\ref{overall_likelihood}, $\langle v\rangle$ is the average
velocity weighted by measurement errors and $\sigma_m$ is the
uncertainty on this weighted average for the combined measurements of
each star.  Note that we assume that the center-of-mass velocity
distribution of Segue~1 is Gaussian.  We also use metallicity ($W'$)
and position to help determine membership, so that the full likelihood
is of the form $\mathcal{L}(v_{i},W',r)$, but we omit the metallicity
and position dependence in the equations presented here for
simplicity.  The absolute magnitude of each star is taken into account
so that its radius can be calculated and only binaries with
separations larger than the stellar radius are allowed, as described
in \citet{minor10} and Paper~II.

For each star, the $J(\mu,\sigma|\mathscr{P})$ factor is generated by
running a Monte Carlo simulation over the distribution of binary
properties, which include the orbital period, mass ratio, and orbital
eccentricity.  Unfortunately, the characteristics of binary
populations in dwarf galaxies are completely unknown at present.  The
best empirical constraints on binary properties come from studies of
the Milky Way, but there is no guarantee that the small-scale star
forming conditions in the Milky Way and those that prevailed in
Segue~1 many Gyr ago are similar, especially since Segue~1 has a
metallicity two orders of magnitude below that of the Galactic disk
population.  In principle, a lower metallicity could change the
probability of forming binary systems (and higher order multiples) and
the properties of those systems, but binaries with separations in the
range that can affect our observations ($\lesssim 10$~AU) are expected
to form via disk fragmentation \citep[e.g.,][]{kratter10} or
interactions between protostellar cores \citep[e.g.,][]{bate09},
neither of which should be very sensitive to metallicity (M. Krumholz
2010, private communication).

In the Monte Carlo simulations, we fix the distributions of the mass
ratio and eccentricity to follow that observed in solar neighborhood
binaries.  However, since the period distributions of binary
populations have been observed to differ dramatically from cluster to
cluster \citep[e.g.,][]{bk98,patience02}, we allow for a range of
period distributions.  Specifically, we assume the distribution of
periods has a log-normal form similar to that of Milky Way field
binaries \citep{dm91,raghavan10}.  We take the mean log period
$\mu_{\log{P}}$ and the width of the period distribution
$\sigma_{\log{P}}$ as free model parameters.  We further assume that
the period distribution observed in Milky Way field binaries is the
result of superposing narrower binary distributions from a variety of
star-forming environments.  Thus, our prior on the period distribution
is that it is narrower than, but consistent with being drawn from, the
\citet{dm91} period distribution with $\mu_{\log{P}} = 2.23$ and
$\sigma_{\log{P}} = 2.3$.  To accomplish this, we choose a flat prior
in $\sigma_{\log{P}}$ over the interval [0.5, 2.3] and a Gaussian
prior in $\mu_{\log{P}}$ centered at $\mu_{\log{P}} = 2.23$, with a
width chosen such that when a large number of period distributions are
drawn from these priors, they combine to reproduce the Milky Way
period distribution of field binaries.  We can then write the
likelihood of each star being a member of Segue~1 or the Milky Way as

\begin{eqnarray}
\lefteqn{\mathcal{L}(v_{i}|e_{i},t_{i},M_{V};f,B,\sigma,\mu,\mathscr{P}) =}  \nonumber \\ 
& & (1-f)\mathcal{L}_{MW}(v_{i}|e_{i}) + 
  f\mathcal{L}_{S1}(v_{i}|e_{i},t_{i},M_{V};B,\sigma,\mu,\mathscr{P}), 
\label{likelihood_of_membership}
\end{eqnarray}

\noindent
where $f$ is the fraction of the total sample that are Segue~1
members.  The second term in Equation~\ref{likelihood_of_membership}
is given by Equation~\ref{overall_likelihood}, and the likelihood
of membership in the Milky Way is

\begin{equation}
\mathcal{L}_{MW}(v_{i}|e_{i}) \propto \int \frac{e^{-\frac{(v_{cm} -
\langle v\rangle)^2}{2\sigma_{m}^2}}}{\sqrt{2\pi\sigma_{m}^2}}
P_{MW}(v_{cm}) dv_{cm}.
\end{equation}

\noindent
The expected Milky Way velocity distribution is taken from a
Besan\c{c}on model \citep{robin03} after the same photometric criteria
used to select Segue~1 stars have been applied.

\subsection{Binary Correction Results}
\label{binary_correction}

\begin{figure*}[th!]
\epsscale{1.17}
\plottwo{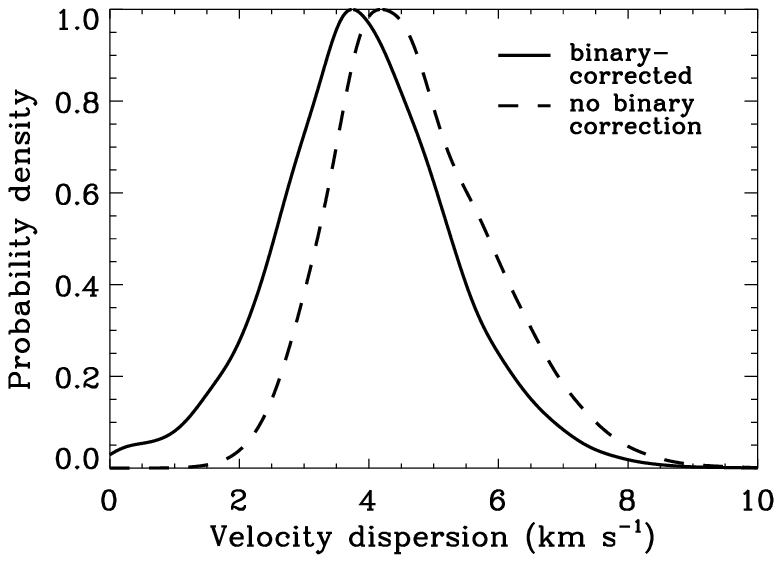}{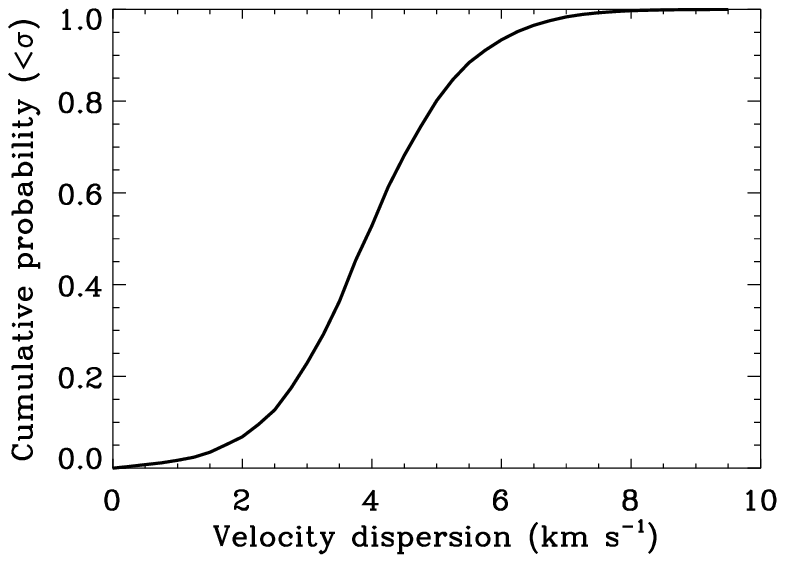}
\caption{(\emph{a}) Posterior probability distribution for Segue~1
  velocity dispersion before (dashed) and after (solid) correcting for
  binary stars.  (\emph{b}) Cumulative probability distribution for
  Segue~1 velocity dispersion after correcting for binaries.  }
\label{dispersion_fig}
\end{figure*}

By applying the above analysis to our full Segue~1 data set we can
correct for the likely presence of binaries in the sample and derive
the intrinsic velocity dispersion of the galaxy.  All observed stars
(not only the likely members) that meet the photometric selection cut
described in \S\,\ref{targetselection} are included in this
calculation, with the weight for each star determined by its
likelihood of membership in Segue~1.  Our posterior probability
distribution for the dispersion is maximized at $\sigma = 3.7$~\kms,
$\sim12$\% smaller than what we measure without a binary correction.
The $1\sigma$ uncertainties on the dispersion are $+1.4$~\kms\ and
$-1.1$~\kms.  We find a 90\% lower limit on the dispersion of
1.8~\kms, and the probability of the true velocity dispersion being
less than 1~\kms\ is $\sim4$\%.  The differential and cumulative
probability distributions for $\sigma$ are displayed in
Figure~\ref{dispersion_fig}.  If the gravitational potential of
Segue~1 were provided only by its stars, the velocity dispersion would
be $\lesssim 0.4$~\kms\ (G09).  \emph{Our lower limit on the dispersion
  therefore allows us to conclude with high confidence that Segue~1 is
  dynamically dominated by dark matter unless it is currently far from
  dynamical equilibrium.}  We discuss the implausibility of large
deviations from equilibrium caused by tidal forces later in
\S\,\ref{tidal_signatures}.

We further note that small intrinsic velocity dispersions ($\sigma
\lesssim 2$~\kms) can only be obtained if the binary period
distribution is skewed toward short periods.  In particular, mean
periods of less than $\sim40$~yr are required to produce such a small
dispersion (Paper~II); for comparison, the mean period in the solar
neighborhood is 180~yr \citep{dm91}.  The posterior distribution we
derive for the mean period in Segue~1 is in fact weighted toward quite
short periods ($\sim10$~yr).  This result is not an artifact of the
short time baseline of our multi-epoch data (observations on
timescales of $\sim1$~yr cannot constrain periods of centuries or
longer), because our priors allow for flatter period distributions
than the posterior for the mean period.  Despite this preference for
shorter periods, the data strongly indicate only a modest contribution
to the velocity dispersion from binaries because the period
distribution is still wide and the small number of detected binaries
indicates that the fraction of stars in close binary systems is not
large.

\begin{figure}[th!]
\epsscale{1.20}
\plotone{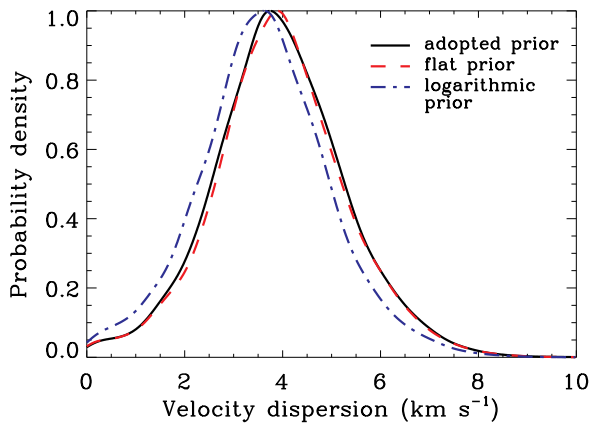}
\caption{Effect of varying the prior distribution for the mean binary
  period on the derived velocity dispersion.  The black solid curve
  shows on the Segue~1 velocity dispersion for our preferred
  assumption of the Milky Way field binary prior from \citet{dm91}.
  The dashed red curve represents a prior that is flat in
  $\mu_{\log{P}}$, and the dash-dot blue curve illustrates the result
  of a logarithmic prior that is even more strongly biased toward
  short periods.  The very small changes in both the most likely value
  of the velocity dispersion and the size of the tail to low values of
  the dispersion ($\sigma \le 1$~\kms) demonstrate that our results
  are robust to differing assumptions about the binary population in
  Segue~1. }
\label{sigma_prior}
\end{figure}

To provide reassurance that our priors on the period distribution are
not biasing the posterior period distribution toward longer periods
(and hence the corrected velocity dispersion toward higher values), we
repeated the calculations above with a flat prior on the mean period
$\mu_{\log{P}}$.  The best-fit mean period barely changed, and while
somewhat shorter periods are allowed in this case, the probability of
an intrinsic velocity dispersion less than 1~\kms\ does not increase
(see Figure~\ref{sigma_prior}).  Even with a more extreme logarithmic
$\mu_{\log{P}}$ prior, the likelihood of a small dispersion is
unchanged despite the resulting very short mean period.  The reason
for this outcome is that when the mean period is forced to be short,
the binary fraction is then constrained to be low and the width of the
period distribution is similarly constrained to be small to fit the
observed changes in the velocities and the observed velocity
distribution.  Hence, the tail of the probability distribution toward
low velocity dispersions cannot be made significantly larger by having
a prior that biases the result to shorter mean periods.

These results contrast with the findings of \citet{mc10}, who conclude
that galaxies like Segue~1 could have very low intrinsic velocity
dispersions that are inflated substantially by the presence of binary
stars.  While we agree with their results given the assumptions they
make, two primary factors are responsible for the different
conclusions from our analysis.  First, \citeauthor{mc10} ignore
binaries with periods longer than $10-100$~yr.  This cutoff appears
reasonable from an observational perspective, since such binaries will
not be detectable in current data sets, but it has the effect of
making the binary fractions they require very large because the
majority of Milky Way binaries have periods longer than 100~yr.
Second, they analyze only single-epoch velocity data sets, whereas the
multiple measurements we have for a number of stars give us much
greater leverage with which to determine the inflation of the velocity
dispersion caused by binaries.  Our Bayesian analysis including the
multi-epoch data and all the information in the tail of the velocity
distribution shows that a substantial inflation by binaries is
disfavored for the Segue~1 data set presented here (see Paper~II for
more details).

\subsection{Mass of Segue~1}
\label{mass}

The same Bayesian machinery described in \S\,\ref{binary_methods} for
determining the velocity dispersion of Segue~1 can also be employed to
calculate the mass of the galaxy (again including a correction for
binary stars).  \citet[][also see \citealt{walker_mhalf}]{joewolf}
derived a simple formula for the mass within the half-light radius of
a system: $M_{1/2} = 3\sigma^{2}r_{1/2}/G$.  For a flat prior on
$\sigma$ (see Paper~II for a discussion of the effect of the choice of
priors), we find a posterior probability distribution on the mass
within the 3D half-light radius of Segue~1 (38~pc) of
$5.8^{+8.2}_{-3.1} \times 10^{5}$~M$_{\odot}$, consistent with the
mass determined by G09 from the original data set.\footnote{Note that
  calculating $M_{1/2}$ directly from the stellar velocity data set
  with this Bayesian approach is not the same as simply plugging the
  derived values of $\sigma$ and $r_{1/2}$ into the
  \citeauthor{joewolf} formula.  The final value for $M_{1/2}$ is very
  similar, but the uncertainties are more accurately determined (in
  particular the 2 and 3$\sigma$ confidence intervals) when we have
  determined the full probability distribution.}  Because the
uncertainties on the mass are not Gaussian, this measurement disagrees
with the stellar mass of Segue~1 ($\sim1000$~M$_{\odot}$) at much more
than $1.8\sigma$ significance; the 99\% confidence lower limit on the
mass is $16000$~M$_{\odot}$ (however, the magnitude of this
low-mass/low-$\sigma$ tail in the probability distribution is
prior-dominated).  The V-band mass-to-light ratio within the
half-light radius is $\sim3400$~M$_{\odot}$/L$_{\odot}$.  Since there
is no evidence that the dark matter halo of Segue~1 is truncated at
such a small radius, this value represents a lower limit on the total
mass-to-light ratio of the galaxy, which could be 1--2 orders of
magnitude larger.

\section{A Stream at 300~\kms}
\label{stream}

In addition to Segue~1 and the Milky Way foreground, we clearly detect
a third population of stars in our kinematic data, at a heliocentric
velocity of 300~\kms.  We refer to this structure as the
``300~\kms\ stream'' because of its lack of spatial concentration
within our survey area.  However, we recognize that these stars could
still be part of a bound system as long as the angular size of the
object is comparable to or larger than our field of view
($\rm{diameter} \gg 20\arcmin$, which corresponds to a physical size
of at least $116~[d/20~\rm{kpc}]$~pc).  As an example, a galaxy
similar to And~XIX, which has a half-light radius of $\sim1.7$~kpc,
would subtend several degrees at this distance \citep{mcconnachie08}.

\begin{figure*}[th!]
\epsscale{1.20}
\plotone{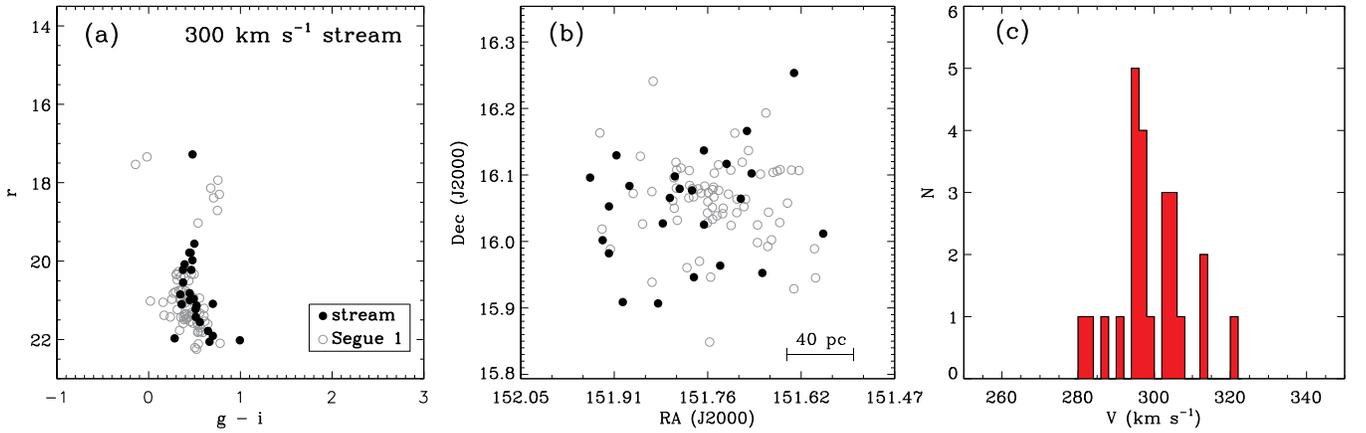}
\caption{(\emph{a}) Color-magnitude diagram of observed stars in the
  300~\kms\ stream.  The filled black circles represent stars
  identified as candidate stream members, while the open gray circles
  are the Segue~1 members.  (\emph{b}) Spatial distribution of
  observed stars in the stream.  Symbols are the same as in
  (\emph{a}).  (\emph{c}) Velocity histogram of
  observed stars in the stream.  Velocities are corrected to the
  heliocentric rest frame, and the velocity bins are 2~\kms\ wide. }
\label{cmd_radec_vhist_stream}
\end{figure*}

G09 also recognized the existence of this stream, finding 4 stars in
it among their smaller sample.  We now present conclusive confirmation
that this structure is real, with $\sim20$ stars in our new data set
(see Figures~\ref{cmd_radec_vhist_fig}--\ref{r_v_ew_mem}).  Because of
the smaller member sample and the low contamination from Milky Way
stars at such extreme velocities, the probabilistic membership
algorithms described in \S\,\ref{selection} are not necessary in this
case.  Instead, we select stars that have velocities between
275~\kms\ and 325~\kms\ (the exact velocity limits are not important;
see Fig.~\ref{cmd_radec_vhist_stream}c) and meet the same photometric
criteria that were used for Segue~1.  The color/magnitude screen
eliminates 5 stars, leaving 24 likely members in the stream.

The CMD of these stars is similar to that of Segue~1, indicating that
the stream is also $\sim20$~kpc away (see
Figure~\ref{cmd_radec_vhist_stream}).  The stream main sequence
appears to be slightly redder (suggesting a higher metallicity) and
slightly closer than Segue~1, although any differences are near the
limit of what can be determined from the SDSS data.  The stream stars
are matched quite well with the fiducial sequence of the globular
cluster M~5 ($\rm{[Fe/H]} = -1.27$) from \citet{an08}, supporting the
higher metallicity that one would have guessed by eye.  By comparing
to various globular cluster sequences, we estimate a distance of
$\approx 22$~kpc and a metallicity of $\rm{[Fe/H]} \approx -1.3$, but
the uncertainties on both numbers are substantial.

The 24 candidate members have a mean velocity of $298.8 \pm
1.7$~\kms\ and a velocity dispersion of $7.0 \pm 1.4$~\kms, comparable
to the dispersion of other known streams
\citep*{chapman08,grillmair08,odenkirchen09,newberg10}.  Several of
these stars may be foreground contaminants; in particular, the bright
star at $g-i = 0.48$, $r = 17.60$ (SDSSJ100720.00+160137.5) is
located a bit below the horizontal branch if the distance of 22~kpc
preferred by the main-sequence fitting is used, although it does still
lie within the AGB selection region shown in
Figure~\ref{cmd_selection}.  However, if the stream is at a slightly
larger distance then this star could well be a horizontal branch
member.  The bluest of the faint stars (SDSSJ100650.83+160351.2; $g-i
= 0.28$, $r = 21.97$) is $\sim3\sigma$ away from the fiducial
sequence used for the original target selection, and two other stars
(SDSSJ100732.48+160500.5 and SDSSJ100708.38+155646.3) have reduced
CaT EWs that are well above those of the bulk of the stream
population.  Even if we remove all four of these stars, the stream
properties do not change significantly; the mean velocity in that case
is $298.7 \pm 1.5$~\kms\ and the dispersion is $5.6 \pm 1.2$~\kms.

If we assume that the 300~\kms\ stream is a bound structure with a
half-light radius larger than our survey area, the \citet{joewolf}
formula implies a lower limit on the mass contained within its
half-light radius of $5.3 \times 10^{6}$~(r/116~\rm{pc})~M$_{\odot}$.
This value would place the stream on the mass-radius relation of Milky
Way dwarfs for a 3D half-light radius of $\sim500$~pc \citep{joewolf}.
On the other hand, it may be worth noting that the stream appears to
be more extended in the east-west direction than north-south
(cf. Figures~\ref{cmd_radec_vhist_fig}\emph{b} and
\ref{cmd_radec_vhist_stream}\emph{b}), consistent with an east-west
extent.  We therefore tentatively suggest that the 300~\kms\ stream
could be the kinematic counterpart of the similarly oriented
photometric feature identified by \citet[][see
  \S\,\ref{no09_stream}]{no09}.  An alternative possibility is that
the stream might be related to Leo~I, which is located approximately
3.8\degr\ due south of Segue~1 at a similar velocity ($282.9 \pm
0.5$~\kms; \citealt*{mateo08}) and metallicity ($\rm{[Fe/H]} = -1.20$;
\citealt{kirby10b}).  Tracing the stream with wider field
spectroscopic data to test these hypotheses would be very desirable.

\section{IS SEGUE 1 UNDERGOING TIDAL DISRUPTION?}
\label{tides}

\citet[][hereafter NO09]{no09} argued that rather than being a bound,
dark matter-dominated dwarf galaxy, Segue~1 is a tiny star cluster
whose apparent velocity dispersion has been inflated by contamination
from the Sagittarius (Sgr) stream.  We have demonstrated that Segue~1
is not a star cluster (\S\,\ref{metallicity}), but that finding does
not address the issues of tidal disruption or contamination.  However,
our new observations and a reanalysis of the SDSS data present some
difficulties for the NO09 hypothesis.  NO09 base their argument on
several key points: (1) over a large area around Segue~1, there are
very low surface density features whose color-magnitude diagrams are
very similar to that of Segue~1 itself; (2) the surface brightness
profile of Segue~1 appears to depart from a standard King or Plummer
model at large radii; (3) the \citet{fellhauer06} model of the Sgr
stream predicts that there should be some very old Sgr material near
the position and velocity of Segue~1; (4) using SDSS blue horizontal
branch (BHB) stars, a coherent feature that can probably be identified
with the Sgr stream approaches the position and velocity of Segue~1;
(5) given the larger velocity dispersion of Sgr, only a small amount
of contamination of the Segue~1 member sample by Sgr stars is
necessary to substantially inflate the apparent velocity dispersion.

Below we discuss each of these ideas in turn and consider how our
results affect their interpretation.  We show that: (1) while
photometric tidal features are indeed present in this field, they
appear more likely to be associated with other kinematically detected
tidal structures rather than Segue~1; (2) in the radial range where
the photometric and kinematic constraints are good, the surface
brightness profile is well described by a Plummer model; (3)
observational evidence for older wraps of the Sgr stream is
non-existent, and even if present, more recent models suggest that
this material has a very low surface density and differs in velocity
from Segue~1; (4) the BHB feature identified by NO09 as potentially
contaminating the Segue~1 data set is offset noticeably in both
position and velocity from Segue~1; (5) the resulting contamination
has therefore been overestimated, and furthermore, these BHB stars
seem to be associated with the Orphan Stream (in which case they are
too spatially confined to affect observations near Segue~1) rather
than the Sgr stream.  We therefore conclude that contamination by Sgr
stream stars does not have a significant impact on the measured
velocity dispersion of Segue~1.  We then consider the evidence that
Segue~1 could be tidally disrupting, finding that while it is not
possible to rule out recent tidal disturbances, the existing data do
not provide significant support for such an interaction.

\subsection{Reconsidering the Disrupting Cluster Scenario}
\label{no}

\subsubsection{Extended Tidal Debris Near Segue~1}
\label{no09_stream}

As NO09 have shown, there is no doubt that there are spatially
extended structures whose stars roughly follow the Segue~1 fiducial
sequence distributed over a wide area around Segue~1.  Within
$\sim1\degr$ of Segue~1, this population is even visible by eye in
SDSS color-magnitude diagrams (CMDs).

What is less obvious is that these stars are necessarily associated
with Segue~1.  One alternative is that they are instead part of the
Sgr stream.  After all, it is clear from both observations
\citep{belokurov07} and models \citep*[G09,][]{law05} that the Sgr
stream passes through this part of the sky at a distance similar to
that of Segue~1.  Indeed, at the position of Segue~1, the stream runs
nearly east-west \citep{belokurov06}, exactly matching the orientation
of the features identified by NO09.  Another possibility is that the
tidal features could be related to the 300~\kms\ stream, which also
shares a very similar stellar population to Segue~1 (note that at the
relevant distances, most of the stars detected by Sloan are on the
main sequence, and thus the CMD filtering is primarily sensitive only
to distance, not to metallicity).  In either case, it seems more
natural to associate this apparent tidal debris with one of the two
known tidal structures at this position, rather than with the one
object that is not obviously undergoing tidal disruption.

\subsubsection{The Surface Brightness Profile of Segue~1}

A second facet of the NO09 picture is the apparent excess of stars
above the fitted Plummer, King, and exponential models at large radii.
However, NO09 themselves note that the area covered by their deeper
imaging is not large enough to define a meaningful background level,
calling into question the significance of this excess.  Within the
radius probed by our kinematic data ($\sim3$ half-light radii), their
photometric analysis shows that the data are fit well by a Plummer
model, in agreement with the distribution of spectroscopically
confirmed member stars that we derive (see
\S\,\ref{tidal_signatures}).  ``Extratidal'' excesses similar to the
one claimed in the outer parts of Segue~1 have been seen in many other
dwarf spheroidals
\citep*{ih95,majewski00,majewski05,md01,palma03,walcher03,wilkinson04,
  mcs05,westfall06,munoz06,sohn07,komiyama07,smolcic07}, but there is
still no consensus regarding their physical significance
\citep{pmn08,penarrubia09}.

\subsubsection{Ancient Wraps of the Sagittarius Stream}
\label{oldwraps}

The Sagittarius dwarf has both leading and trailing tidal streams
stretching across the entire sky.  The most recent wrap of the streams
has been detected robustly in numerous ways
\citep[e.g.,][]{ibata01,vivas01,dohm-palmer01,bellazzini03,newberg03,majewski03,belokurov06}.
Models predict that Sgr debris stripped on previous orbits may be
present as well, but there is currently little observational evidence
for such material.  G09 showed that while Segue~1 is spatially
coincident with the leading arm of the Sgr stream, the velocities of
the recently stripped stars differ from that of Segue~1 by
$\sim100$~\kms, firmly ruling out an association.  NO09 pointed out
that the \citet{fellhauer06} model predicts that there are also Sgr
stars stripped several orbits earlier at this position that have
velocities similar to Segue~1.  In the most recent model by
\citet{lm10a}, which is the most successful to date in matching
observations,\footnote{While the \citet{lm10a} model provides a
  generally reasonable match to Sgr stream data, the bifurcation of
  the stream in the SDSS footprint is not yet fully understood and is
  not present in the model.  \citet{penarrubia10} argue that this
  structure can be a result of the original internal kinematics of the
  Sgr dwarf if its progenitor was a disk galaxy.} however, stars in
this ancient wrap uniformly have much lower velocities in this part of
the sky ($v_{GSR} < 6$~\kms, compared to $v_{GSR} = 113$~\kms\ for
Segue~1).  In addition, the surface density associated with the stars
stripped at the earliest times should be overwhelmingly smaller than
that of the more recent debris.  Within 10\degr\ of Segue~1, the
\citet{lm10a} simulation contains $\gtrsim 200$ times as many stars in
the recently-stripped, leading (negative velocity) stream as are
present in the older, trailing stream.  Since NO09 find similar
numbers of stars in their observed positive and negative velocity BHB
streams, this provides a strong argument that the BHB structure closer
in velocity to Segue~1 is not, in fact, related to Sgr, as we discuss
further in \S\,\ref{contamination}.  Without any quantitative evidence
for significant numbers of Sgr stream stars that are close in both
position and velocity to Segue~1, we do not see any reason to expect
substantial Sgr contamination of our Segue~1 member sample.

\subsubsection{Does the Sagittarius Stream Overlap in Velocity With Segue~1?}
\label{overlap}

NO09 also used SDSS observations of BHB stars to trace the kinematics
of the \emph{observed} Sgr debris near Segue~1 (while plenty of
observations of Sgr stream velocities exist in other parts of the sky,
the velocities near Segue~1 had not previously been measured).  They
found that the main component of the stream has negative heliocentric
(and galactocentric) velocities at this position, more than
200~\kms\ away from the velocity of Segue~1, as pointed out by G09.
Another coherent BHB component, though, is present at much higher
velocities, similar to the prediction from \citet{fellhauer06} for
older Sgr debris.  NO09 concluded from this result that there is
likely confusion between Segue~1 stars and Sgr stream stars in both
position and velocity.  However, \emph{even accepting for the moment
  that these stars are actually part of the Sgr stream} (see
\S\S\,\ref{oldwraps} and \ref{contamination} for our
counterarguments), two factors significantly diminish this confusion.
First, the BHB stream identified by NO09 does not actually appear to
reach the location of Segue~1; it peaks at $\delta \approx +24\degr$,
$\sim8\degr$ north of Segue~1, and seems to have petered out by the
time it reaches Segue~1.  Equally important is that the velocity of
the BHB stars is $V_{hel} \approx 195$~\kms\ ($V_{GSR} \approx
132$~\kms), offset from the velocity of Segue~1 by 13~\kms\ in the
heliocentric frame and 19~\kms\ in the Galactic Standard of Rest (GSR)
system.

\subsubsection{Contamination of the Segue~1 Member Sample}
\label{contamination}

Relying on the line of reasoning examined above, NO09 proposed that
Segue~1 is actually a star cluster whose derived properties have been
distorted by contamination from Sgr stream stars.  Such contamination
is a difficult issue to quantify, because by definition any stars that
could be contaminating the member sample must have very similar
velocities, metallicities, distances, and ages to Segue~1 stars.  The
only way to assess definitively the expected number of contaminants
would be with an even wider field survey to identify Segue~1-like
stars that are far enough away from the galaxy so as to be very
unlikely to be associated.  While SDSS includes some of the desired
data, the SDSS spectroscopic coverage of stars at faint magnitudes is
very sparse, so the vast majority of stars do not have velocity
measurements.  Nevertheless, some do, and those observations can be
used to estimate the significance of the contamination.

The NO09 estimate of the contamination depends critically on the
assumptions discussed in \S\,\ref{overlap} that the BHB stream they
identify with Sagittarius is exactly coincident in both position and
velocity with Segue~1.  However, as noted in \S\,\ref{overlap}, the
surface density of BHB stars in the higher velocity component appears
to be down by a factor of at least a few by the time it reaches
Segue~1 (their Fig. 10), and the velocity offset compared to Segue~1
further reduces the contribution of these stars within the Segue~1
velocity selection window.

To take into account these effects, we repeat the analysis described
by NO09.  Using DR7 data, if we select the BHB stars according to
their heliocentric velocities, the positive velocity stream component
has a mean velocity of $v_{hel} = 195$~\kms.  Since the stream is
extended spatially, and therefore likely has a velocity gradient along
its length, it is more concentrated in the GSR velocity system, where
its velocity is $v_{GSR} = 132$~\kms\ and its velocity dispersion is
12.2~\kms\ (corresponding to an intrinsic dispersion of 10~\kms\ after
the 7~\kms\ median velocity errors are removed).  Given the velocity
window spanned by the likely Segue~1 members (not including
SDSSJ100704.35+160459.4) of $194.6~\mbox{\kms} \le v_{hel} \le
224.2$~\kms, only 41\% of stars in the positive velocity stream would
be expected to have velocities consistent with membership in Segue~1.
Applying our actual photometric selection criteria
(\S\,\ref{targetselection}) instead of the broader selection box used
by NO09 reduces our estimate of the surface density of Sgr stream
stars at the declination of Segue~1 to 120~deg$^{-2}$ (for stars with
r magnitudes between 17.5 and 21.7).  The complete region of our
spectroscopic survey covers 0.087~deg$^{2}$, and the effective area of
the full survey (including the incompleteness at larger radii) is
$\sim0.14$~deg$^{2}$.  Assuming, as NO09 did, that half of these stars
are in the negative velocity stream component, and removing the 59\%
of the positive velocity stream stars that would still lie outside the
Segue~1 velocity range, suggests that a total of $2-3$ Sgr stream
stars could be in our sample.

The next question is what effect including a few Sgr stars in an
analysis of Segue~1 would have.  We repeat the Monte Carlo simulation
carried out by NO09 to answer this question.  Using the same setup
they did, with assumed velocity dispersions of 1~\kms\ for Segue~1
and 10~\kms\ for the Sgr stream (and putting them both at the same
mean velocity, contrary to the argument above), we find that 5 Sgr
stars must be included in the 71 star Segue~1 sample to have a
significant chance ($\sim20$\%) of boosting the apparent velocity
dispersion of Segue~1 to at least 3.9~\kms.  Given the smaller number
of contaminating stars estimated above, we conclude that the inclusion
of Sgr stream stars in the Segue~1 member sample is not likely to
provide the dominant component of the observed velocity dispersion.

\begin{figure}[t!]
\epsscale{1.24}
\plotone{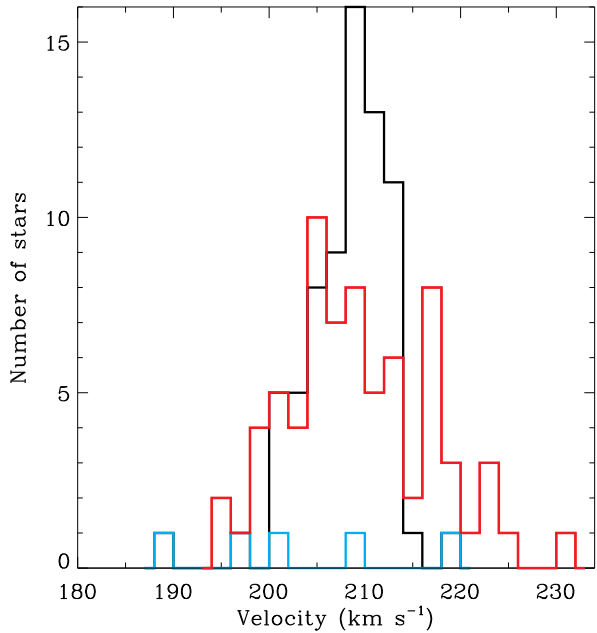}
\caption{Monte Carlo simulation of the expected velocity distribution
  that would be obtained in the presence of significant contamination
  by the Sgr stream.  We assume intrinsic dispersions of 1~\kms\ for
  Segue~1, 10~\kms\ for the Sgr stream, mean velocities of
  208~\kms\ for both components, median velocity errors of
  5~\kms\ (with a minimum of 2.2~\kms), and 5 Sgr stars in a 71 star
  sample.  The velocities of the full simulated sample are shown in
  black, with the Sgr contaminants overplotted in cyan and the
  observed Segue~1 stars in red.}
\label{sgr_sim}
\end{figure}

Also, when significant numbers of such contaminants are present they
tend to have an easily visible effect on the velocity distribution (as
they must if they are going to alter the dispersion).  Visually, the
simulated velocity histograms frequently appear to be composed of a
narrow central peak containing most of the stars, surrounded by a few
well-separated outliers (see Fig. \ref{sgr_sim} for an example).
These outliers would raise suspicions in any membership classification
scheme like the ones outlined in \S\,\ref{selection}, and might well
be discarded from the sample.  Interlopers that happen to fall within
the main peak of the velocity distribution (and are therefore more
difficult to identify) do not have a significant impact on the
velocity dispersion; only stars in the wings of the distribution can
both be mistaken for members and substantially change the apparent
dispersion.  However, our analysis shows that any such contaminating
population cannot be large, and in Paper~II we demonstrate that
including an additional population does not change the derived
parameters for Segue~1.

\begin{figure}[t!]
\epsscale{1.17}
\plotone{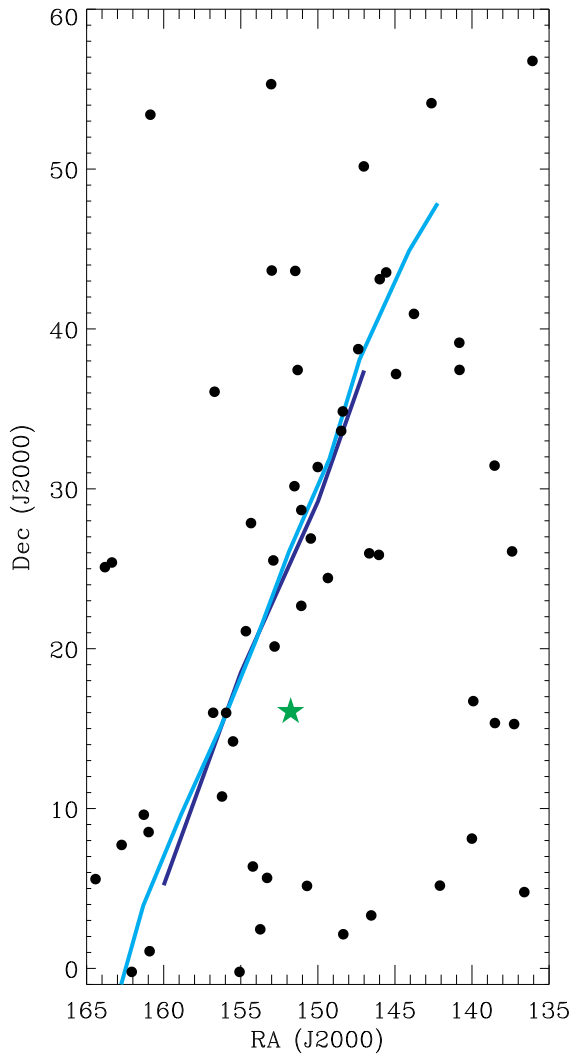}
\caption{Spatial distribution of the BHB stars that can be confidently
  associated with the positive velocity stream identified by NO09.
  The green star marks the position of Segue~1, and the cyan and blue
  lines indicate the traces of the Orphan Stream from
  \citet{newberg10} and \citet{orphan_stream}, respectively.  The close
  correspondence between the path of the Orphan Stream and the
  positions of the BHB stars at the same velocity suggests that these
  stars are members of the Orphan Stream rather than the Sagittarius
  stream.}
\label{orphan_stream}
\end{figure}

Moreover, a closer examination of the positive velocity stream
component calls into question the assumption that it is associated
with Sagittarius at all.  \citet{newberg10} used BHB stars in SDSS and
the SEGUE survey to trace the Orphan Stream across the sky and noted
that it passes slightly north of Segue~1, at a distance of
$\sim25$~kpc and a velocity of $v_{GSR} = 130$~\kms.  Isolating the
SDSS BHB stars within 1~$\sigma$ of the positive velocity stream's
mean velocity, we find that their spatial distribution closely matches
the track of the Orphan Stream determined by \citet{orphan_stream} and
\citet{newberg10}, as shown in Figure~\ref{orphan_stream}.  The good
correspondence between the path of the Orphan Stream and the BHB stars
at the same velocity is highly suggestive that these stars are members
of the Orphan Stream rather than an old (and heretofore undetected)
wrap of the Sagittarius stream.  If this peak is indeed related to the
Orphan Stream, which is narrow and spatially confined, and not
Sagittarius, then it is quite unlikely that any main sequence stars
associated with this feature would be located close enough to Segue~1
to appear as contaminants in our survey.  That would then imply that
most or all of the Sgr stars near Segue~1 are at negative velocities,
as suggested in \S\,\ref{oldwraps}, which would further reduce our
estimate of the possible contamination by Sgr above.

After considering each piece of evidence in concert, we thus conclude
that contamination by the Sgr stream does not provide a very plausible
explanation for the large velocity dispersion of Segue~1.

\subsection{Signatures of Tidal Disruption}
\label{tidal_signatures}

Having determined the nature of Segue~1, the effect of binary stars on
the velocity dispersion, and the level of contamination by the Sgr
stream, the final issue we must analyze is the impact of Milky Way
tides.  Unfortunately, while the presence of tidal tails would be
incontrovertible evidence of tidal effects, the contrapositive is not
true: there are no observations that can conclusively rule out tidal
disruption.  We therefore consider several possible signatures of
tides.

\begin{figure}[t!]
\epsscale{1.24}
\plotone{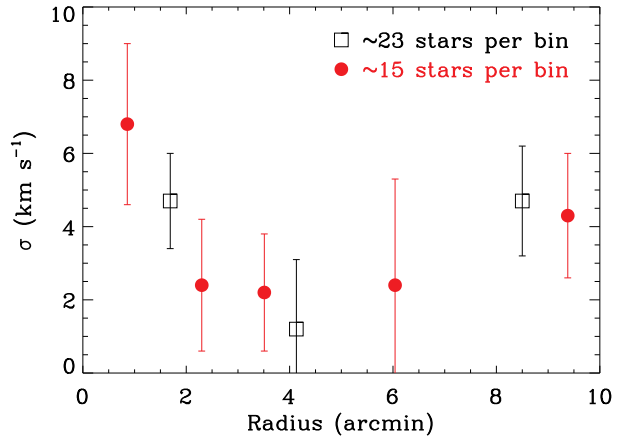}
\caption{Velocity dispersion profile of Segue~1.  The open squares
  show the profile obtained for bins of $\sim23$ stars each, and the
  filled red circles show the profile for bins of $\sim15$ stars.
  While the decrease in the velocity dispersion at intermediate radii
  (also visible in Figure~\ref{r_v_ew}) does not appear to be an
  artifact of the binning, it is only significant at the $1~\sigma$
  level.}
\label{dispersion_prof}
\end{figure}

\begin{itemize}

\item First, our spectroscopic survey shows there are no obvious tidal
  tails connected to Segue~1 (see Figure~\ref{cmd_radec_vhist_fig}).
  Although the spatial distribution is not uniform, we find Segue~1
  members in every direction around the galaxy rather than the bipolar
  pattern that tidal tails would be expected to produce.  The apparent
  clumpiness of the member stars toward the western edge of the galaxy
  may simply be the result of small number statistics
  \citep{martin08}.  While the spectroscopic member sample confirms
  that Segue~1 has an elliptical shape, nonzero ellipticities are not
  necessarily associated with tidal influences \citep*{mmj08} and may
  just reflect the shape with which the galaxy formed.  We also note
  that the tidal tails seen in the SDSS photometry in the Segue~1
  discovery paper \citep{belokurov07} have not been confirmed by
  deeper followup (\citealt{belokurov07}, NO09, Mu{\~n}oz et al., in
  prep.).

\item Velocity gradients are a commonly used indicator of tidal
  disruption, although like tidal tails they may only be visible in
  particular geometries and at large radii \citep{pp95,mmj08,lokas08}.
  We see no velocity gradient across Segue~1; the mean velocities of
  the stars in the eastern and western halves of the galaxy agree
  within their uncertainties.  The inclusion or exclusion of the
  ambiguous members, RR~Lyraes, and binaries does not affect this
  result.  There are also no apparent trends in the mean velocity from
  one end of the galaxy to the other \citep[unlike, e.g.,
    Willman~1;][]{willman10}.  Using the methodology described by
  \citet{strigari10} to calculate completely general constraints on
  the rotation of Segue~1, we derive a 90\% confidence upper limit of
  5.0~\kms\ on the rotation amplitude, but we note that samples a
  factor of 2 larger than ours are typically required to detect
  rotation.

\item A velocity dispersion profile that rises at large radii is often
  regarded as a possible result of tidal stripping, although the same
  behavior can be interpreted as evidence for an extended dark matter
  halo as well.  Our member sample is not large enough to divide the
  data into more than a few radial bins, but we can begin to
  investigate the shape of the dispersion profile.  Velocity
  dispersion profiles with two different binnings are displayed in
  Figure~\ref{dispersion_prof}.  As seen in the left panel of
  Figure~\ref{r_v_ew}, the velocity dispersion appears to reach a
  local minimum at a radius of $\sim3$\arcmin\ before increasing back
  to its central value at larger radii.  This shape seems to be
  independent of the exact binning chosen (and in fact is visible in
  the unbinned data), but as the error bars in the figure show, it is
  not statistically significant.  While it is possible that the
  dispersion increase could suggest that the stars beyond
  $\sim8$\arcmin\ from the center of the galaxy have been stripped, we
  caution that apparent features in other data sets of similar size
  (or even larger) have often disappeared when larger samples of
  velocity measurements become available \citep{wilkinson04,kleyna04}.
  With the modest number of bins possible for a sample of 71 members,
  we view the shape of the dispersion profile of Segue~1 as possibly
  interesting but not necessarily meaningful at this point.  It is
  also worth pointing out that the mass implied by the central
  velocity dispersion, \emph{even if the decline at $\sim3$\arcmin\ is
    real}, is enough to put the tidal radius beyond the observed
  extent of the galaxy (see below), suggesting a consistency problem
  for the tidal interpretation.

\item Finally, ``extratidal'' excesses of stars at large radii are
  frequently considered to be indicative of tidal disturbances.  The
  \citet{king62} tidal radius or limiting radius of Segue~1 is not
  well known because of the lack of deep enough wide-field photometry
  (although NO09 estimate a value of $\sim26$\arcmin), but our
  complete spectroscopic sample enables us to investigate the stellar
  profile out to $r \sim 13$\arcmin.  Our observations are effectively
  complete within 2 half-light radii of the center of the galaxy
  (\S\,\ref{completeness}), where we identify 61 member stars.  Between
  2 and 3 times the half-light radius we obtained successful spectra
  for 62 out of 75 stars located in the highest priority photometric
  selection region, for a completeness of 83\%.  We therefore adjust
  the 9 observed members in that annulus to a projected total of 11
  members, yielding 72 member stars within 3 half-light radii
  (13.2\arcmin).  We find 39 member stars within $1~r_{\rm{half}}$
  (54\% of the total), 22 member stars between $1~r_{\rm{half}}$ and
  $2~r_{\rm{half}}$ (31\%), and estimate 11 member stars between
  $2~r_{\rm{half}}$ and $3~r_{\rm{half}}$ (15\%), compared to the
  expected numbers of 56\%, 33\%, and 11\% for a Plummer profile (once
  the 10\% of the stars that should lie beyond $3~r_{\rm{half}}$ are
  removed from consideration).  The radial profile of Segue~1 thus
  does not show any excess out to at least $3~r_{\rm{half}}$ (88~pc).

\end{itemize}

The above arguments demonstrate an absence of evidence in favor of
tidal disruption, but as mentioned at the beginning of this section
and discussed in detail by \citet{mmj08}, none of them (singly or in
concert) are sufficient to prove that Segue~1 is not being tidally
disrupted.  Perhaps the strongest evidence for the absence of tidal
effects results from consistency checks between the mass we measure
for Segue~1, the corresponding tidal radius, and the timescale for
tidal disruption.

As discussed by G09, the current position of Segue~1 is difficult to
reconcile with a scenario in which it is in the final stages of
disruption.  Segue~1 has a crossing time of $\sim10^{7}$~yr, and at a
velocity of $\sim200$~\kms\ it will travel less than 2~kpc per
crossing time.  In order for the observed kinematics of Segue~1 to be
significantly distorted by tides, the galaxy must be within a few
crossing times of its orbital pericenter.  Conservatively, then,
Segue~1 should be no more than $\sim10$~kpc past pericenter.  Segue~1
is located 28~kpc from the Galactic Center, which would place its
pericenter at a Galactocentric distance of at least 18~kpc,
inconsistent with the closest approach to the Milky Way that would be
required to disrupt it (see below).

For the IAU value of the Milky Way rotation velocity (220~\kms), the
mass enclosed at the position of Segue~1 is $3 \times
10^{11}$~M$_{\odot}$.  The mass of Segue~1 is best constrained at the
half-light radius of the galaxy \citep{joewolf}, where we obtain
$M_{\rm{half}} = 5.8^{+8.2}_{-3.1} \times 10^{5}$~M$_{\odot}$.  Even
if we assume (without any physical basis) that the mass distribution
is arbitrarily truncated at the half-light radius, the instantaneous
Jacobi radius \citep[e.g.,][Equation 8.91]{bt08} for Segue~1 would be
$\sim250$~pc.  If we use the Jacobi radius as an estimate of the tidal
radius, then all of the stars we observed are well within the
present-day tidal radius.  Although \citeauthor{bt08} present a
detailed discussion of why the Jacobi radius is necessarily an
imperfect estimator, it is worth noting that subhalos in the Aquarius
simulation \citep{springel08} show a strong correlation between the
radius of the subhalo and the Jacobi radius, supporting the use of the
Jacobi radius as the tidal radius in practice.  In order to bring the
tidal radius in to the position of our outermost confirmed member, the
pericenter of Segue~1's orbit must be less than $\sim10$~kpc.
Substantially disturbing stars \emph{at the half-light radius}, where
the mass is being measured, requires an orbital pericenter of less
than $\sim4$~kpc (eccentricity greater than 0.75, since Segue~1 is
clearly not near its apocenter at the present time).  The Jacobi
radius scales as $M_{\rm{Segue~1}}^{1/3}$, so even major revisions to
the derived mass do not affect our conclusion.  This calculation is
quite conservative, because it relies most strongly on the central
kinematics of the galaxy where the observational uncertainties are
smallest and tidal effects are weakest.  

Incorporating reasonable assumptions about the extent of the dark
matter halo of Segue~1 only strengthens this result.  In the Via
Lactea II simulation \citep{vl2}, subhalos with $V_{\rm max} >
10$~\kms\ (see \S~\ref{darkmatter}) that currently reside between 20
and 40~kpc from the host halo have median tidal truncation radii of
$\sim500$~pc.  Since these simulations self-consistently include
tides and orbital trajectories, there is good reason to suspect that
the mass of Segue~1 extends well past $r_{\rm{half}}$.  If we
extrapolate the Segue~1 mass beyond the observed region using CDM
priors (since current simulations cannot resolve radii smaller than
$\sim 100$~pc), we find a mass within 100~pc of $M_{100} = 2.2 \times
10^{6}$~M$_{\odot}$ and a mass within 300~pc of $M_{300} = 1.4 \times
10^{7}$~M$_{\odot}$, consistent with the common mass scale of Milky
Way satellites \citep{strigari08a}.  With these larger masses, the
current tidal radius would increase to $400-700$~pc, making the center
of Segue~1 nearly impervious to tides for any plausible orbit.

\section{THE IMPORTANCE OF SEGUE~1 FOR DARK MATTER STUDIES}
\label{darkmatter}

The large estimated mass of Segue~1 and its very small size (it has
the smallest half-light radius of any known Local Group dwarf galaxy,
with the possible exception of Willman~1) mean that Segue~1 also has
the densest known concentration of dark matter.  The average density
enclosed within its half-light radius is
$2.5^{+4.1}_{-1.9}$~M$_{\odot}$~pc$^{-3}$, substantially higher than
that found in other dwarf galaxies
(\citealt{gilmore07,sg07,walker_mhalf,tollerud11} [Fig. 17]) or the
solar neighborhood \citep[e.g.,][]{bahcall84}.  For comparison
purposes, this density is equal to the ambient density of dark matter
at $z\simeq 300$ and the average density of objects that collapsed at
$z\simeq 50$.  Given the extremely high mass-to-light ratio of Segue~1
(\S\,\ref{mass}), it is safe to equate the dark matter density with
the total density.

In the context of $\Lambda$CDM, densities as high as those that we
infer within the half-light radius of Segue~1 are indicative of {\em
  massive} subhalos.  Using the mass estimator from \citet{joewolf},
the circular velocity at the half-light radius is related to the
measured velocity dispersion via $V(r_{{\rm half}}) = \sqrt{3} \,
\sigma \simeq 6.4$~\kms.  This demands that $V_{\rm max} >
6.4$~\kms\ for the halo hosting Segue~1.  Convolving the central
circular velocity with CDM-based priors suggests that $V_{\rm max} >
10$~\kms\ \citep[e.g.,][]{bullock10}.  Such halos are indeed found in
the most advanced current N-body simulations \citep{vl2,springel08},
demonstrating that the density of Segue~1 is reasonable in a
$\Lambda$CDM universe, but how common it is for galaxies of Segue~1's
luminosity to be found in very massive subhalos is not yet clear.

Since the flux of high-energy particles from dark matter annihilation
scales as $\rho_{{\rm DM}}^{2}r^{3}/d^{2}$ and Segue~1 is also the
second-nearest dwarf galaxy to the Sun, Segue~1 is clearly a high
priority target for indirect detection experiments
\citep{martinez09,essig09,scott10}.  \citet{essig10} use the sample
of member stars presented here to carry out more detailed calculations
of the expected gamma-ray and neutrino flux from dark matter
annihilation in Segue~1.  That analysis shows that Segue~1 is expected
to be among the two brightest sources of annihilation radiation from
Milky Way satellites, and may be the brightest known dwarf galaxy.  We
strongly encourage future indirect detection searches for dark matter
to target Segue~1.

Finally, the high density of Segue~1 provides important leverage for
constraints on the phase-space density of dark matter particles, which
is often estimated by the related quantity $Q_{{\rm DM}}(r) =
\rho_{{\rm DM}}(r)/\sigma_{{\rm DM}}(r)^{3}$ defined by \citet{hd00}.
Unfortunately, $Q_{{\rm DM}}$ cannot be measured directly from
velocity dispersion data.  While $\rho_{{\rm DM}}$ may be determined
fairly accurately within $r_{\rm half}$, $\sigma_{\rm DM}$ is not
observable.  Generally, we expect $\sigma_{\rm DM} > \sigma_* $
\citep[e.g.,][]{joewolf} because the dark matter velocity dispersion
is governed by the total mass beyond the stellar radius (which cannot
be measured).  This implies that $\rho_{\rm DM}/ \sigma^{3}_{*}$
provides only an {\em upper limit} on $Q_{\rm DM}$ at any particular
radius.  Some caution is advisable when reading the literature on this
subject.

Given that $Q_{\rm DM}$ cannot be measured directly, we must rely on
model fitting inspired by a theory prior in order to quantify $Q_{\rm
  DM}$ constraints from Segue~1.  Of particular interest is the case
of warm dark matter (WDM), where the primordial phase-space density
may produce observationally accessible cores in the dark matter
density.  It is therefore useful to assume a dark matter density
profile that is compatible with WDM rather than the usual CDM
\citet*{nfw} or similar profile. We use a cored isothermal profile for
illustration, which should have approximately the right shape for such
models.  With this profile and the kinematic data presented in this
paper, we determine the posterior probability density for $Q_{\rm DM}$
at equally spaced logarithmic radii out to the stellar tidal radius
following, e.g., \citet{strigari08b}.  We assume uniform priors on the
scale radius and scale density for the isothermal profile, and we
additionally make the assumption of velocity isotropy for the stars
and the dark matter.  Under these assumptions, we find a lower limit
on the central value for $Q_{\rm DM}$ to be
$\sim10^{-3}$~M$_{\odot}$~pc$^{-3}$~(\kms)$^{-3}$, higher than that
determined for any other galaxy.\footnote{SG07 list somewhat higher
  values for a few systems, but those were derived under much
  different assumptions (most notably, that mass follows light).  For
  a consistent set of assumptions, Segue~1 has a higher phase-space
  density than any of the dwarfs analyzed there.}  Estimates of this
kind can place a lower limit on the allowed mass range for various WDM
candidates, and we suggest that more detailed treatments of the
phase-space density in Segue~1 to quantify these constraints would be
very worthwhile.

\section{SUMMARY AND CONCLUSIONS}
\label{conclusions}

We have presented a comprehensive spectroscopic survey of the
ultra-faint Milky Way satellite Segue~1.  The observations were
designed both to search for potential tidal debris around Segue~1 and
to constrain the effects of binary stars on its velocity dispersion.
Within a radius of 10\arcmin\ (67~pc) from the center of Segue~1, we
measured the velocities of 98.2\% of the candidate member stars.  We
identified 71 likely members, which we used to study the metallicity,
kinematics, and nature of Segue~1.

The six red giants in Segue~1 have a mean metallicity of $\rm{[Fe/H]}
= -2.5$, and span a range of nearly 2~dex from $\rm{[Fe/H]} = -3.4$
to $\rm{[Fe/H]} = -1.6$.  Both the presence of extremely metal-poor
stars with $\rm{[Fe/H]} < -3$ and the enormous metallicity spread
demonstrate unambiguously that Segue~1 is a galaxy, rather than a
globular cluster as some previous studies have suggested.

Using the new Bayesian method presented in our companion paper
\citep{martinez}, we analyzed the kinematics of the entire observed
data set, allowing for contamination by Milky Way foreground stars and
employing repeated velocity measurements for a subsample of the
targets to correct for the effect of binary stars.  We derived an
intrinsic velocity dispersion of $3.7^{+1.4}_{-1.1}$~\kms\ for Segue~1
and only a 2\% probability that the dispersion is small enough to be
provided by the stellar mass of Segue~1 alone.  The estimated mass
contained within the half-light radius is $5.8^{+8.2}_{-3.1} \times
10^{5}$~M$_{\odot}$, giving Segue~1 a V-band mass-to-light ratio at
that radius of $\sim3400$~M$_{\odot}$/L$_{\odot}$.

Based on updated data and models, we re-examined earlier proposals
that Segue~1 is tidally disrupting and that kinematic studies of it
are likely to be contaminated by the Sagittarius stream.  We showed
that there is no observational evidence supporting the possibility of
tidal disruption, and that the tidal radius of Segue~1 has likely
always exceeded its stellar extent unless it has an orbital pericenter
around the Milky Way of less than $\sim4$~kpc.  We also determined
that contamination by Sgr stream stars is significantly lower than
previously estimated; our current member sample is unlikely to contain
more than 3 contaminants, which is not enough to substantially inflate
the velocity dispersion.

Taken together, the results of our observations clearly point to the
interpretation that Segue~1 is a dark matter-dominated galaxy --- in
fact, it has the highest mass-to-light ratio, and is therefore the
darkest galaxy, yet found.  The mean density inferred for Segue~1
within its half-light radius is consistent with the extrapolated
density profiles of massive subhalos in high resolution $\Lambda$CDM
galactic halo simulations \citep*{mdk08,springel08}.  The relative
proximity of Segue~1 makes a strong case for considering Segue~1 in
future searches for the products of dark matter annihilation processes
\citep[e.g.,][]{essig10}.  The density of dark matter within the inner
38~pc of Segue~1, $2.5^{+4.1}_{-1.9}$~M$_{\odot}$~pc$^{-3}$ or
$\sim100$~GeV/c$^{2}$~cm$^{-3}$, is the highest dark matter density yet
determined, and consequently has broad implications for particle
physics models and galaxy formation on small scales.

\acknowledgements{JDS gratefully acknowledges the support of a Vera
  Rubin Fellowship provided by the Carnegie Institution of Washington.
  MG acknowledges support from NSF grant AST-0908752.  Work at UCI was
  supported by NSF grant PHY-0855462 and NASA grant NNX09AD09G.
  Support for this work was also provided by NASA through Hubble
  Fellowship grant HST-HF-01233.01 awarded to ENK by the Space
  Telescope Science Institute, which is operated by the Association of
  Universities for Research in Astronomy, Inc., for NASA, under
  contract NAS 5-26555.  BW acknowledges support from NSF grant
  AST-0908193.  We thank Vasily Belokurov, Michael Cooper, Gerry
  Gilmore, Juna Kollmeier, Mark Krumholz, David Law, and George
  Preston for helpful conversations, and Matt Walker for providing his
  EM code.  We also thank Alan McConnachie and Pat C{\^o}t{\'e} for
  sharing a draft of their work on binary stars prior to publication.
  The analysis pipeline used to reduce the DEIMOS data was developed
  at UC Berkeley with support from NSF grant AST-0071048.  This
  research has also made use of NASA's Astrophysics Data System
  Bibliographic Services.}

{\it Facilities:} \facility{Keck:II (DEIMOS)}

\clearpage
\begin{deluxetable*}{lllcccccccc}
\tablecaption{Segue~1 Velocity Measurements}
\tablewidth{0pt}
\tablehead{
\colhead{Star} &
\colhead{$\phn$Velocity} &
\colhead{$\phn\phn\Sigma$Ca} &
\colhead{Radius} &
\colhead{MJD} &
\colhead{g} &
\colhead{r} &
\colhead{i} &
\colhead{Member} &
\colhead{EM Member} &
\colhead{Bayesian} \\
\colhead{} &
\colhead{$\phn$(\kms)} &
\colhead{$\phn\phn$(\AA)} &
\colhead{(arcmin)} &
\colhead{} &
\colhead{} &
\colhead{} &
\colhead{} &
\colhead{(Subjective)\tablenotemark{a}} &
\colhead{Prob.\tablenotemark{b}}  &
\colhead{Member Prob.\tablenotemark{c}} }
\startdata
SDSSJ100613.98+155436.1  &         \phn$-46.8 \pm 4.6$ &       \phs\phn$3.3 \pm 0.6$  &    15.4 &  54890.243 &  24.28 &  22.64 &  21.17 &  0 &  $-9.999$  &  $-9.999$  \\
SDSSJ100614.24+160424.7  &         \phn$-66.7 \pm 2.4$ &       \phs\phn$2.4 \pm 0.3$  &    11.8 &  54890.243 &  20.29 &  19.97 &  19.81 &  0 &  \phs0.000 &  \phs0.000 \\
SDSSJ100614.31+160050.9  &         \phs$150.1 \pm 3.3$ &       \phs\phn$2.9 \pm 0.5$  &    12.3 &  54890.243 &  21.52 &  21.03 &  21.13 &  0 &  \phs0.000 &  \phs0.000 \\
SDSSJ100614.40+160013.0  &      \phs\phn$70.1 \pm 2.2$ &       \phs\phn$4.8 \pm 0.3$  &    12.5 &  54890.243 &  17.14 &  16.44 &  16.16 &  0 &  \phs0.000 &  \phs0.000 \\
SDSSJ100614.78+155512.7  &      \phs\phn$45.2 \pm 2.9$ &       \phs\phn$1.5 \pm 0.4$  &    14.8 &  54890.243 &  24.83 &  22.50 &  21.06 &  0 &  $-9.999$  &  $-9.999$  \\
SDSSJ100614.87+160858.7  &      \phs\phn$29.1 \pm 2.9$ &       \phs\phn$4.4 \pm 0.4$  &    12.5 &  54890.243 &  20.09 &  19.55 &  19.32 &  0 &  \phs0.000 &  \phs0.000 \\
SDSSJ100615.16+155556.6  &      \phs\phn$54.3 \pm 2.6$ &       \phs\phn$2.9 \pm 0.3$  &    14.3 &  54890.243 &  22.30 &  21.07 &  19.96 &  0 &  $-9.999$  &  $-9.999$  \\
SDSSJ100615.53+160056.9  &      \phn\phn$-6.7 \pm 2.7$ &       \phs\phn$1.7 \pm 0.4$  &    12.0 &  54890.243 &  21.04 &  20.73 &  20.65 &  0 &  \phs0.000 &  \phs0.000 \\
SDSSJ100616.95+160524.3  &         \phn$-10.0 \pm 2.4$ &       \phs\phn$3.3 \pm 0.3$  &    11.2 &  54890.243 &  23.22 &  21.47 &  20.39 &  0 &  $-9.999$  &  $-9.999$  \\
SDSSJ100617.35+155606.6  &   \phs\phn\phn$1.7 \pm 2.4$ &       \phs\phn$2.1 \pm 0.4$  &    13.8 &  54890.243 &  22.81 &  21.47 &  20.68 &  0 &  $-9.999$  &  $-9.999$  \\
\enddata
\label{datatable}
\tablecomments{This table will be published in its entirety in the
  electronic edition of the \emph{Journal}.  A portion is shown here
  for guidance regarding form and content.}
\tablenotetext{a}{Member status according to the criteria established
  at the beginning of \S~\ref{selection}.  1 indicates membership, and
  0 is for non-members.}
\tablenotetext{b}{Membership probability from the EM algorithm.
  The algorithm is run on the subset of stars whose
  colors and magnitudes are consistent with membership, so photometric
  non-members are indicated by probability $-9.999$.}
\tablenotetext{c}{Membership probability from the Bayesian approach.
  As with the EM algorithm, these calculations are run on the subset
  of stars whose colors and magnitudes are consistent with membership,
  so photometric non-members are indicated by probability $-9.999$.}
\end{deluxetable*} 
\clearpage

\begin{deluxetable*}{lcccccccc}
\tablecaption{Segue~1 Metallicity Measurements}
\tablewidth{0pt}
\tablehead{
\colhead{Star} &
\colhead{g} &
\colhead{r} &
\colhead{i} &
\colhead{$T_{\rm{eff}}$} &
\colhead{$\log{g}$} &
\colhead{[Fe/H]} &
\colhead{Number of} &
\colhead{Dispersion between} \\
\colhead{} &
\colhead{} &
\colhead{} &
\colhead{} &
\colhead{} &
\colhead{} &
\colhead{} &
\colhead{measurements} &
\colhead{measurements} }
\startdata
SDSSJ100702.46+155055.3     & 18.48 & 17.94 & 17.73 & $5148 \pm 102$ & $2.64$ & $-2.48 \pm 0.15$ & 2 & 0.42~dex \\
SDSSJ100714.58+160154.5     & 18.83 & 18.30 & 18.06 & $5102 \pm 109$ & $2.78$ & $-1.73 \pm 0.14$ & 3 & 0.07~dex \\ 
SDSSJ100652.33+160235.8     & 18.87 & 18.39 & 18.16 & $5271 \pm 132$ & $2.85$ & $-3.40 \pm 0.17$ & 3 & 0.32~dex \\
SDSSJ100742.72+160106.9     & 19.72 & 18.59 & 18.14 & $5251 \pm 111$ & $2.76$ & $-2.50 \pm 0.14$ & 1 &  \\
SDSSJ100710.08+160623.9     & 19.20 & 18.71 & 18.45 & $5106 \pm 109$ & $2.96$ & $-1.63 \pm 0.14$ & 4 & 0.03~dex \\
SDSSJ100639.33+160008.9     & 19.44 & 19.03 & 18.90 & $5643 \pm 186$ & $3.21$ & $<-3.4$ & 3 &  
\enddata
\label{metallicity_table}
\end{deluxetable*}

\clearpage

\appendix
\section{APPENDIX: COMMENTS ON THE MEMBERSHIP OR NON-MEMBERSHIP OF 
INDIVIDUAL STARS}
\label{appendix_comments}
\medskip

In this appendix, we briefly discuss the case for membership or
non-membership for several stars whose membership status is not
clear-cut.  Unlike SDSSJ100704.35+160459.4, however, the decision of
whether to include or exclude these stars has no appreciable effect on
the derived properties of Segue~1.  We remind the reader that in the
Bayesian approach used for our main results all of these stars (except
the last two) are included in the calculations, weighted according to
their membership probabilities.  Other schemes generally require each
star to be classified as either a member or non-member.

\begin{itemize}

\item \emph{SDSSJ100637.49+161155.1} --- Very much like
  SDSSJ100704.35+160459.4, SDSSJ100637.49+161155.1 sits perfectly on
  the Segue~1 fiducial sequence but has a significantly higher
  velocity (more than $3\sigma$ even for the largest possible value
  for the dispersion) than the galaxy.  Given its relatively large
  distance from the center of Segue~1 (9.7\arcmin) and its higher than
  average CaT EW ($W' = 3.66$~\AA), we consider this star unlikely to
  be a member.  Its EM membership probability is 0.69 and its
  membership probability from \citet{martinez} is 0.23.

\item \emph{SDSSJ100622.85+155643.0} --- SDSSJ100622.85+155643.0 has a
  very large uncertainty on its velocity measurement ($v = 223.9 \pm
  37.8$~\kms) because of the low S/N of its spectrum and an apparently
  broad \ha\ line.  Since much of the $1\sigma$ range for its velocity
  puts the star at a velocity consistent with Segue~1, we subjectively
  classified it as a member, but the statistical algorithms give it
  lower probabilities ($p = 0.73$ for EM and $p = 0.50$ for
  \citealt{martinez}.  However, the large velocity error means that
  SDSSJ100622.85+155643.0 receives essentially no weight in
  determining the Segue~1 velocity dispersion, so its true membership
  status is not important.

\item \emph{SDSSJ100711.80+160630.4} --- SDSSJ100711.80+160630.4 lies
  well outside the Segue~1 velocity range at $v = 247.1$~\kms, but
  also has a large velocity uncertainty of $15.9$~\kms.  The very high
  velocity led us to classify it as a non-member, but the objective
  techniques recognize that there is a reasonable chance that the
  star's actual velocity could be significantly lower, which might
  make it a member.  Weighting its position near the center of Segue~1
  quite heavily, the EM algorithm gives a membership probability of $p
  = 0.98$, while the \citeauthor{martinez} algorithm more
  conservatively estimates $p = 0.70$.  Again, the large velocity
  error minimizes its impact on the derived velocity dispersion.

\item \emph{SDSSJ100743.55+160947.2} --- SDSSJ100743.55+160947.2 has a
  velocity near the high end of the Segue~1 velocity range ($v = 223.4
  \pm 5.3$~\kms), but the uncertainty is large enough for it
  plausibly to be a member.  In the color-magnitude diagram, it is
  located just outside the highest priority selection region, but
  again close enough that the photometric errors certainly allow it to
  be a member.  We consider this star to be a probable member, but the
  objective algorithms give it moderate membership probabilities ($p =
  0.87$ for EM and $p = 0.54$ for the Bayesian determination) because
  of its high velocity and large radius.

\item \emph{SDSSJ100630.96+155543.6} --- SDSSJ100630.96+155543.6 has a
  velocity several standard deviations smaller than the systemic
  velocity of Segue~1, although the large velocity uncertainties (8.7
  and 11.8~\kms\ for the two measurements) make this difference of
  marginal significance.  Combined with its position 11.6\arcmin\ from
  the center of the galaxy, the \citeauthor{martinez} algorithm hedges
  its bets at $p = 0.60$, while the EM membership probability is 0.94.
  This star is another probable member.

\item \emph{SDSSJ100658.11+160701.4} --- With a velocity of $187.0 \pm
  2.3$~\kms\ and a $g-i$ color $\sim0.25$~mag blueward of the red
  giant branch (outside both the high and low priority CMD selection
  regions), SDSSJ100658.11+160701.4 appears to be a clear non-member
  star.  However, the star could conceivably be an SX~Phoenicis
  variable, in which case membership in Segue~1 would be possible,
  although as a variable star it would still be excluded from our
  analysis of the kinematics.  SDSSJ100658.11+160701.4 is redder than
  would generally be expected for an SX~Phoenicis star
  \citep[e.g.,][]{olech05,moretti09}, but we cannot rule out such a
  classification with the available data.

\item \emph{SDSSJ100700.75+160300.5} --- In our standard reduction of
  the data, using the arc spectrum obtained closest to the time of the
  observations, we were not able to determine a reliable velocity for
  SDSSJ100700.75+160300.5.  The cross-correlation with the
  best-fitting template spectrum produced multiple widely separated
  cross-correlation peaks, with no obvious way to identify the correct
  solution.  Switching to the arc frame from the end of the night
  ($\sim10$ hours after the mask was observed) instead of the one from
  the afternoon ($\sim3$ hours before the observations) produced a
  cleaner spectrum with a velocity of $v = 208.4 \pm 9.7$~\kms.
  Because none of the other spectra on this mask required such special
  treatment, and the increased time between calibrations and
  observations provides more opportunity for changes in the
  instrument, we regard this measurement as somewhat questionable and
  omit the star from our sample.  It is probably a member of Segue~1,
  but with a velocity consistent with the systemic velocity of the
  galaxy and a large velocity error, including it would not change any
  of our results.

\end{itemize}

\end{document}